%% file: paper.tex
\documentclass[sigconf]{acmart}
\renewcommand\footnotetextcopyrightpermission[1]{}
\input{preamble}

\begin{document}

\title{Rethinking Diversity in Deep Neural Network Testing}

\author{Zi Wang}
\affiliation{%
  \institution{University of Wisconsin-Madison}
  \city{Madison}
  \country{USA}
}
\email{zw@cs.wisc.edu}

\author{Jihye Choi}
\affiliation{%
  \institution{University of Wisconsin-Madison}
  \city{Madison}
  \country{USA}
}
\email{jihye@cs.wisc.edu}

\author{Ke Wang}
\affiliation{%
  \institution{Visa Research}
  \city{Palo Alto}
  \country{USA}
}
\email{kewang@visa.com}

\author{Somesh Jha}
\affiliation{%
  \institution{University of Wisconsin-Madison}
  \city{Madison}
  \country{USA}
}
\email{jha@cs.wisc.edu}


\input{abstract}


\settopmatter{printacmref=false}



\maketitle

\input{intro}
\input{02_prelim_jc}

\input{03_main}
\input{eval}
\input{conclusion}

\bibliographystyle{ACM-Reference-Format}
\bibliography{paper}
 
\input{appendix}
\end{document}

%% file: preamble.tex
\usepackage{cite}
\usepackage{amsmath,amsfonts}
\usepackage{algorithmic}
\usepackage{graphicx}
\usepackage{textcomp}
\usepackage{xcolor,colortbl}
\def\BibTeX{{\rm B\kern-.05em{\sc i\kern-.025em b}\kern-.08em
    T\kern-.1667em\lower.7ex\hbox{E}\kern-.125emX}}

\usepackage[skins,xparse]{tcolorbox}

\usepackage[utf8]{inputenc} 
\usepackage[T1]{fontenc}    
\usepackage{hyperref}       
\usepackage{url}            
\usepackage{booktabs}       
\usepackage{nicefrac}       
\usepackage{multirow}
\usepackage[capitalize,noabbrev]{cleveref}
\usepackage{rqbox}
\usepackage{supertabular}
\usepackage{microtype}
\usepackage{caption}
\usepackage{subcaption}
\usepackage{adjustbox}
\usepackage{amsthm}
\usepackage{mathtools}
\usepackage{tabularx}
\usepackage{bm}
\usepackage[shortlabels]{enumitem}
\usepackage{wrapfig}

\newcommand{\R}{\mathbb{R}}

\newcommand{\norm}[1]{||#1||}

\newcommand{\func}{\circ}

\newcommand{\act}{\sigma}

\newcommand{\sign}{\text{sign}}

\newcommand{\relu}{\text{ReLU}}

\newcommand{\class}{\mathcal{C}}

\def \ie {{\em i.e.},~}

\DeclareMathOperator*{\argmax}{arg\,max}

\theoremstyle{plain}
\newtheorem{theorem}{Theorem}[section]
\newtheorem{proposition}[theorem]{Proposition}

\theoremstyle{definition}

\theoremstyle{remark}
\newtheorem{remark}[theorem]{Remark}

\renewcommand{\paragraph}[1]{\vspace{.05in}\noindent\textbf{{#1}.~~}}

\definecolor{blush}{rgb}{0.65, 0.16, 0.16}

\NewTotalTColorBox{\mybox}{ O{} m +m }{%
  enhanced,sidebyside,size=tight,boxrule=0pt,frame hidden,
  lower separated=false,fontlower=\sffamily,
  extrude left by=\dimexpr\hoffset+\oddsidemargin+1in\relax,
  extrude right by=\dimexpr\paperwidth-\textwidth-\hoffset-\oddsidemargin-1in\relax,
  colback=gray!30,
  lefthand width=0.2\textwidth,
  sidebyside gap=1cm,
  before skip=8pt,
  after skip=8pt,
  #1
}{\includegraphics[width=\linewidth]{#2}\tcblower #3}

\usepackage{tikz}
\usepackage{pgfplots}
\usepackage{environ}
\usepackage{tikz-cd}
\usetikzlibrary{decorations.pathreplacing}

\pgfplotsset{compat=newest}

\AtEndEnvironment{example}{\hfill$\blacksquare$}%
\AtEndEnvironment{definition}{\hfill$\blacksquare$}%

\pgfkeys{/pgfplots/scale/.style={
  x post scale=#1,
  y post scale=#1,
  z post scale=#1}
}

\pgfplotsset{every tick label/.append style={font=\tiny}}

\usetikzlibrary{arrows,backgrounds,decorations.markings,hobby}
\usepgflibrary{shapes.multipart}
\usetikzlibrary{patterns}
\makeatletter
\newcommand{\pgfplotsdrawaxis}{\pgfplots@draw@axis}
\makeatother

\pgfplotsset{only axis on top/.style={axis on top=false, after end axis/.code={
             \pgfplotsset{axis line style=opaque, ticklabel style=opaque, tick style={thick,opaque},
                          grid=none}\pgfplotsdrawaxis}}}
\pgfplotsset{every axis/.style={scale only axis}}
\tikzset{>=latex} 

\tikzstyle{oper}=[rounded corners, draw=black, thick, minimum size = 4mm]
\tikzstyle{input}=[rounded corners, draw=white, thick, minimum size = 4mm]
\tikzstyle{output}=[rounded corners, thick, draw=white, minimum size = 4mm]
\tikzstyle{empty}=[circle, draw=white, minimum size = 4mm]
\makeatletter
\newsavebox{\measure@tikzpicture}
\NewEnviron{scaletikzpicturetowidth}[1]{%
  \def\tikz@width{#1}%
  \begin{lrbox}{\measure@tikzpicture}%
  \BODY
  \end{lrbox}%
  \pgfmathparse{#1/\wd\measure@tikzpicture}%
  \BODY
}
\makeatother
\newcommand{\true}{standard testing}

\usepackage{array}
\newcolumntype{L}[1]{>{\raggedright\let\newline\\\arraybackslash\hspace{0pt}}m{#1}}
\newcolumntype{C}[1]{>{\centering\let\newline\\\arraybackslash\hspace{0pt}}m{#1}}
\newcolumntype{R}[1]{>{\raggedleft\let\newline\\\arraybackslash\hspace{0pt}}m{#1}}

%% file: abstract.tex
\begin{abstract}

Motivated by the success of traditional software testing, numerous diversity measures have been proposed for testing deep neural networks (DNNs).
In this study, we propose a shift in perspective, advocating for the consideration of DNN testing as directed testing problems rather than diversity-based testing tasks. 
We note that the objective of testing DNNs is specific and well-defined: identifying inputs that lead to misclassifications. Consequently, a more precise testing approach is to prioritize inputs with a higher potential to induce misclassifications, as opposed to emphasizing inputs that enhance "diversity."

We derive six directed metrics for DNN testing. 
Furthermore, we conduct a careful analysis of the appropriate scope for each metric, as applying metrics beyond their intended scope could significantly diminish their effectiveness.
Our evaluation demonstrates that (1) diversity metrics are particularly weak indicators for identifying buggy inputs resulting from small input perturbations,
and (2) our directed metrics consistently outperform diversity metrics in revealing erroneous behaviors of DNNs across all scenarios.
\end{abstract}


%% file: intro.tex
\section{Introduction}
Deep learning (DL) systems have exhibited remarkable performance across a spectrum of machine learning tasks, reshaping modern society~\citep{Krizhevsky_imagenetclassification,nlp_2013}. 
Despite their widespread application, these systems still maintain a largely opaque decision process. To exacerbate the situation, they are shown susceptible to \textit{functionality bugs}~\citep{szegedy2014intriguing,goodfellow2015explaining,debug}. Here, we loosely define functionality bugs as misalignment between the ground-truth of inputs and the DNN classification results. To effectively identify such functionality bugs, this work introduces a new methodology for testing DNNs, in particular, we focus on DNN classification models as the programs under testing, 
given that the DNN classification problem represents a canonical task in deep learning~\citep{deep-learning}. 
In recent years, researchers have proposed various methods to test DNNs~\citep{diversity-boosting,ATS,deepxplore,NC-meaningful}. Many of them are motivated by traditional software testing practices, particularly with an emphasis on improving input diversity. 
The underlying assumption is that diverse inputs can traverse the program more thoroughly, uncovering corner cases~\citep{afl,fuzzing-cacm}.

Contrary to existing work of DNN testing, our testing framework does \emph{not} employ the diversity heuristics. Instead, we advocate for treating DNN testing tasks in the literature~\citep{deepxplore,diversity-boosting} as \emph{directed} testing tasks.  Directed testing is primarily concerned with the detection of specific bugs (\cref{sec:direct}). This approach is better suited to the nature of DNN testing, considering that the functionality of DNNs is clear-cut --- making predictions that align with human visual perception. Therefore, testing DNNs comes with a well-defined, specific objective --- finding inputs that cause DNNs to mispredict (\cref{sec:def}). Take image classification models as an example; in technical terms, this objective entails seeking inputs that lead DNNs to produce values in their logit layer such that the index of the maximum value no longer corresponds to the label of an input image. Given the precision and clarity of this testing goal, it is more appropriate to adopt a directed testing approach that prioritizes inputs that have a higher potential to trigger misclassifications by DNNs, rather than inputs that improve ``diversity''~\citep{aflgo,tofu}.

\paragraph{Inputs Transformations}
Due to the substantial cost involved in obtaining and manually labeling test inputs, we opt to generate inputs through visual-semantics preserving transformations. These transformations augment existing inputs with variations such as unnoticeable noise, image rotation, shifts, and shearing~\citep{szegedy2014intriguing,ATS,shorten2019survey}~(\cref{sec:dnn-test}). 
Regrettably, recent diversity-based testing works overlooked the case small noise can be applied to the inputs~\citep{diversity-boosting,ATS}. This scenario is necessary to consider because minor distortions can arise in data acquisition systems and remain imperceptible to humans. Furthermore, these subtle distortions can lead to misclassifications by DNNs, a phenomenon widely recognized as \emph{adversarial examples}~\citep{szegedy2014intriguing}. Adversarial examples have emerged as a prominent research topic in the field of deep learning over the past few years, inspiring many influential works, including generative adversarial networks and stable diffusion models~\citep{goodfellow2015explaining,gan,diffusion}. Consequently, testing methodologies that overlook adversarial examples risk compromising their credibility.

In contrast, our testing framework can address not only natural input transformations such as image rotation and shifts~\citep{ATS,shorten2019survey} but also small noise. 
In fact, our formal analysis shows that the two types of transformations have significantly different 
mathematical properties despite being 
under the umbrella term of visual semantics preserving transformations.
This distinction has a further impact on the scope of testing methods. 

Moreover, we propose that testing can be expedited by leveraging an intrinsic nature of DL systems: differentiability.
The differentiability of DNNs enables us to derive fast and precise linear approximations of testing metrics 
by utilizing gradients. 
As we explain later (\cref{sec:diff}), adversarial examples created by projected gradient descent (PGD)~\citep{pgd}, state-of-the-art $\ell_p$-adversarial attacks, can be instantiated from our testing framework. In essence, the results of existing adversarial attacks can all be viewed as specific instances of the testing inputs to our framework when we only consider small $\ell_p$-distortions as the input transformations. 

We evaluate our claims over four datasets and three types of standard DNN architectures, in particular, we compare our framework with several diversity-based methods on both standard and metamorphic testing tasks. Our evaluation shows that when input transformation is restricted to small distortions only, our directed metrics are significantly more effective than diversity metrics in identifying inputs that induce functionality bugs in DNNs. More specifically, given the same set of test inputs, those that are highly ranked by our directed metrics consistently lead DNNs to misclassifications. In contrast, inputs that are prioritized by diversity metrics barely trigger the erroneous behavior from DNNs. Moreover, when the other type of input transformations --- natural transformation --- is taken into account,
our directed metrics are again more effective than diversity metrics in exposing the functionality bugs of DNNs.

The core contribution of our work lies in uncovering the shortcomings of diversity metrics and highlighting the advantages of directed metrics for testing DNNs. Furthermore, we demonstrate how these directed testing metrics can be harnessed to effectively detect functionality bugs in the DNN, similar to~\citet{diversity-boosting}. It is worth noting that while our metrics may be applied to train more robust DNNs, this is not the primary focus of our work. Instead, our emphasis is to devise a general and effective framework for testing DNNs based on directed metrics.



To summarize, we make the following contributions:
\begin{itemize}
\setlength\itemsep{0.2em}
    \item Novelty: We provide a novel conceptual framework for DNN functionality testing, and demonstrate that some state-of-the-art DL practices are the natural consequences of this framework; while~\citet{NC-meaningful,nc-person} only refuted the application of neuron coverage, a specific diversity measure. Hence, we are more constructive compared to~\citet{NC-meaningful,nc-person}. 
    \item Formality: We formalize the testing problem and carefully analyze the input transformations. In particular, we demonstrate that although small distortions and natural image transformations do not alter human judgment, they have distinct mathematical properties and should be addressed differently for testing purposes. We justify this distinction both theoretically and empirically. 
    To the best of our knowledge, ours is the first work that has explicitly studied this distinction, and we assert that this nuanced examination is imperative.
    
    \item Verifiability: We evaluate our claims over four datasets and three types of standard DNN architectures, and open source the code and dataset needed for the evaluation. 
\end{itemize}

%% file: 02_prelim_jc.tex
\section{background}
\subsection{Mathematical Notations and Definitions}\label{sec:def}
Let $[n] = \{1,\ldots,n\}$. 
For a vector $v\in\R^n$, $\sign(v) \in \{-1, 0, 1\}^n$ such that 
\[ \sign(v)_i = \left\{ \begin{array}{ll}
            1, & \mbox{$v_i > 0$}\\
            0, & \mbox{$v_i = 0$}\\
            -1, & \mbox{$v_i < 0$}\end{array} \right. 
\]
Let $\norm{v}_p$ denote the $\ell_p$ norm of a vector $v$, \ie
\[\norm{v}_p = \sqrt[\leftroot{-1}\uproot{8}\scriptstyle p]{\sum_{i=1}^n |v_i|^p}.\]
The canonical Euclidean norm is then $\norm{v}_2$. Another commonly considered norm on the input is the $\ell_\infty$-norm, defined as 
\[
\norm{v}_\infty = \max_{i\in[n]}{|v_i|}.
\]

For two functions $f$ and $g$, $f\func g(x) = f(g(x))$ denotes the composition of $f$ and $g$. If $f:\R^n\rightarrow \R$ is differentiable, we use $D^{i}f$ to denote the $i$-th order differentiation of partial derivatives, e.g., $D^{1}f$ is the gradient of $f$, and $D^{2}f$ is the Hessian of $f$. As a convention, we omit $1$ when we denote gradient, i.e., we use $Df$ to denote the gradient of $f$.

For an infinitely differentiable multivariate function $f:\R^n\rightarrow \R$, the Taylor expansion $T$ of $f$ at $a\in \R^n$ is
\begin{equation}\label{eq:taylor}
T(a+\Delta x) = f(a)+(\Delta x)^TDf(a)+\frac{1}{2}(\Delta x)^TD^2f(a) (\Delta x)+\ldots,
\end{equation}
Therefore, $T(a+\Delta x)$ is an infinite series. If $f$ is analytic, then $T(a+\Delta x)$ converges to $f(a+\Delta x)$~\citep{babyrudin}. In this work, we only work with the first two terms of $T$; that is, we only consider 
\begin{equation}\label{eq:linear}
    f(a)+(\Delta x)^TDf(a).
\end{equation}

Given the information of $f$ at $a$ (\ie $f(a)$ and $Df(a)$), ~\cref{eq:linear} is the linear approximation of $f$ at $a+\Delta x$. This can provide an estimation of $f(a+\Delta x)$ if we do not want to evaluate $f$ at $a+\Delta x$ directly: $f(\Delta x+a)\approx f(a)+(\Delta x)^TDf(a)$. In the supplementary materials, we illustrate a linear-approximation example in the one-dimensional case (\cref{fig:linear_approx}).

\begin{remark}\label{rm:quality}
The quality of linear approximation depends on how large $\Delta x$ is, and how close $f$ is to a linear function. The smaller $\Delta x$ is, the higher precision that approximation is, the closer the result of~\cref{eq:linear} is to~\cref{eq:taylor}. If $f$ is already linear, then all the higher-order beyond linear terms become $0$, and~\cref{eq:taylor} degenerates to~\cref{eq:linear}.
\end{remark}

\paragraph{DNNs}
We hereby discuss some basics of DNNs that are needed to elaborate our testing approach. 
A feed-forward DNN $f:\R^m\rightarrow \R^l$ is a composition of affine transformations and non-linear activation functions:
\[
f_1(x)=  W^{(1)} x+b^{(1)};\; f_i(x)=  W^{(i)}\act(x)+b^{(i)}, i=2,\ldots,d.
\]
where $W^{(i)}\in \R^{n_{i+1}\times n_{i}}$ is the weight matrix between the layers with $n_1 = m$ and $n_{d+1}=l$, $d$ is the depth of the network, and $b^{(i)}\in \R^{n_{i+1}}$ is the bias term. $\act$, the activation, is an element-wise non-linear function. Also, we have $f = f_d\func\cdots\func f_1$. Notice that practical activation functions are (almost) everywhere infinitely differentiable except at one point (e.g., $x=0$ for $\relu$). However, this does not impact the validity of the linear approximation. If $f$ is infinitely differentiable at $x$ and $\Delta x$ is small, then the $f$ is also infinitely differentiable at the neighborhood centered at $x$.
\begin{remark}\label{rm:non-linear}
    In general, a neural network is highly non-linear. Therefore, when $f$ in~\cref{eq:linear} is realized as a DNN, for the linear approximation to be precise, it is necessary for $\Delta x$ to be small.
\end{remark}

In this work, our focus is on DNNs for standard classification tasks. 
$f: \R^m\rightarrow \R^l$ has $l$ outputs, and let $f^{(i)}$ be the $i$-th output of $f$. The classification of an input $x$ is $\class(f,x)=\argmax_{k\in[l]}f^{(k)}(x)$. Suppose that the prediction of $x$ is $i$, then $f^{(i)}(x)> f^{(j)}(x)$ holds for all $j\neq i$. The output of $f$ is called the logit score, and the classification model outputs the class with the highest logit score.

\subsection{Software Testing}
\label{sec:st}
Large software artifacts have been an integral part of modern society for decades and are evolving on a constant basis. Despite its ubiquity, software usually contains various defects, and some of them can raise serious security concerns. Testing remains one of the most efficient ways to demonstrate the presence of bugs. Since modern software is complicated, bugs can occur anywhere in the program. For instance, a buffer usage in the program could potentially introduce overflow. 

\paragraph{Diversity in Software Testing}
In general, automated testing techniques are commonly guided by diversity heuristics: a diverse set of inputs can execute the program thoroughly and therefore, is likely to expose more bugs~\citep{fuzzing-cacm}. One of the diversity measures in testing is code coverage with multiple granularities encompassing function coverage, basic block coverage, and path coverage. 
For instance, American Fuzzy Lop (AFL), one of the most popular fuzzing tools, is a state-of-the-art \textit{general-purpose} fuzzer that employs the coverage maximization heuristics~\citep{afl}. It repeats a loop involving input
mutation, execution of programs, and evaluation of coverage-based diversity metric. AFL generates inputs aiming to execute different parts of the program and has successfully detected numerous bugs in various applications.

\paragraph{Directed Software Testing}\label{sec:direct}
Unlike general-purpose software testing tools such as AFL, directed testing prioritizes inputs that are close to the \textit{specific} testing goal. 
For instance, a data-race testing tool should prioritize inputs that are more likely to trigger concurrency bugs rather than inputs executing unseen parts of programs~\citep{Razzer}. 
In software testing, this proximity can often be measured statically by distance in code. Specifically, inputs that trigger data-races must traverse code that can be executed concurrently, so the proximity can be measured by the distance between the current execution trace and target race conditions in the program. Accordingly, the testing objective should be mutating inputs to lead executions 
moving closer to the targets and eventually triggering the specific bugs~\citep{tofu}. 

Researchers have developed several directed testing tools, and shown that they are more capable of detecting specific bugs than general-purpose testing tools~\citep{aflgo}. 
Inspired by these findings, in the following~\cref{sec:dnn-test}, we illustrate that the objective of DNN testing is well-defined and specific; thereby, DNN testing should conceptually be considered as directed testing.



\subsection{DNN Testing}\label{sec:dnn-test}
\paragraph{Diversity in DNN Testing}\label{sec:diverse-dnn-testing}
Inspired by the success of traditional software testing, the previous literature on DNN testing has been focused on devising various diversity measures. One of the first diversity measures is neuron coverage (NC)~\citep{deepxplore}, similar to the code coverage in traditional software testing, which measures how many neurons are activated above a specified threshold $k$. The assumption is that the more neurons are activated, the more states of DNNs are explored. Later works refined the definition of NC and proposed several structural variants of NC, including the neuron boundary coverage and strong neuron activations~\citep{deep_gauge}. 
More recently,~\citet{diversity-boosting} proposed to use intermediate layer activation divergence between metamorphic inputs as an indicator to detect prediction violations, arguing that inputs with different intermediate activation patterns should also have divergent predictions.~\citet{ATS} proposed a measure to compute the fault pattern of inputs based on the normalized logit score, and a test selection algorithm that chooses a subset of inputs with diverse fault patterns and higher prediction uncertainty.

\paragraph{Testing objective}
The functionality of DNNs is to classify inputs correctly aligned with human perception. Accordingly, functionality bug in the classification setting is characterized as incorrect classification, and the testing goal is to generate inputs that lead to misclassification. 
However, obtaining such test inputs and manually labeling them is usually a very expensive process. 
To address this test generation problem, metamorphic testing~\citep{metamorphic} has been proposed to augment existing inputs with metamorphic relations. 
Since the functionality of vision models is to align the model judgment with human perception, one important metamorphic relation is \emph{equivalence-preserving} image transformations under the visual semantics.
We refer to these transformations as visual semantics-preserving transformations.
Now, instead of manually collecting and labeling new test inputs, one could simply augment the original dataset by applying a few compositions of such visual semantics-preserving transformations on labeled inputs, and labels are supposed to be retained since the semantics of inputs are preserved.
If any of those transformed inputs alter the DNN's prediction, then we identify successful test inputs for \emph{functionality testing} (\ie test inputs that trigger misclassification).

\paragraph{Visual semantics-preserving transformations}
There exist two types of transformations that leave human judgment on input semantics unaffected. The first type comprises \emph{natural transformations}, such as image rotation, zooming, and brightness adjustments. The second type involves \emph{small distortions} constrained by the $\ell_p$-norm, which is popular under the name of adversarial examples in DL literature~\citep{szegedy2014intriguing,goodfellow2015explaining}. 
Despite this, recent prominent works on diversity-based DNN testing have predominantly focused on natural transformations~\citep{diversity-boosting}, even explicitly overlooking small distortions~\citep{ATS}. 
In this paper, we aim to propose a functionality testing framework that effectively incorporates both types of perturbations. This is achieved by approaching DNN testing as directed testing rather than diversity-based testing.

\subsection{Our Main Thesis}

\paragraph{Simple semantics of DNN}
The core message of this paper is to highlight that diversity-based methods inspired by software testing are not the ideal testing framework for DNN testing focused on functionality bugs.
Compared with traditional software, which might contain complicated structures like pointer aliasing, various data structures, and indirect function calls,
DNNs are specialized numerical programs only involving a few operations: matrix addition and multiplication, activation function, and the max function and vector indexing. 
Hence, the techniques designed to handle the complexity in traditional software might not be directly applicable to DNNs. 
Such observation naturally leads us to the main thesis of our work:
\begin{focusbox}
Functionality testing of DNNs is a directed testing task. 
\end{focusbox}
This is because, under the simple semantics of DNN, the mathematical description of the testing objective (\ie goal of functionality testing as described earlier in~\cref{sec:dnn-test}) can be very specified, which will be elaborated in~\cref{sec:thesis}.
Furthermore, our perspective based on directed testing toward DNN testing enables addressing various input transformations (both natural transformations and small distortions), while the perspective of previous works based on diversity metrics is only effective with natural transformations.

\begin{remark}
  Even if small distortions and natural image transformations are under the same umbrella term, visual semantics-preserving transformations, they are mathematically distinct transformations. Natural image transformations like image rotations can change the pixel representation of images significantly. 
\end{remark}

%% file: 03_main.tex
\section{proposed metrics for DNN testing}
\label{sec:thesis}
We first combine the DNN semantics and functionality testing goal to derive the forward fitness metrics, which are suitable for natural transformations~(\cref{sec:forward}). 
Then, we apply the linear approximation in~\cref{eq:linear} to the forward fitness metrics to derive the backward counterparts using the gradients, suitable for small perturbations on images~(\cref{sec:diff}). 
Finally, we combine both forward and backward scores to address the case where both natural input transformations and small perturbations are allowed~(\cref{sec:mixed}). 
The lesson is that the effectiveness of the fitness metrics depends on the underlying input transformations, so one needs to carefully analyze the transformations before using the metrics. 


\subsection{Forward Fitness}\label{sec:forward}
We use $x$ to denote a human-labeled input, and let $x'$ be a generated input from a few small visual semantics-preserving transformations, similar to input mutation in traditional software testing. We fix a DNN under test: $f$. Let $i\in [l]$ be $x$'s label. We assume that $f$ classifies $x$ correctly, i.e., $C(f, x)= i$. 
Clearly, the goal of functionality testing is to identify the buggy $x'$ so that the $f$'s classification over $x'$ will change, i.e., $C(f, x')\neq C(f, x)$. We provide two different interpretations of bugs, each of which can provide a testing metric.


\paragraph{Margin score}
$C(f, x)= i$ implies $f^{(i)}(x) > f^{(j)}(x), \forall j\in[l], j\neq i$. We define 
\[g_{ji} = f^{(j)}-f^{(i)}.\] 
where $g_{ji}(x)<0$, $\forall j\neq i$, because $C(f, x)= i$.
To change the prediction of $f$ on $x'$ that is close to $x$ semantically, we need $f^{(j)}(x') > f^{(i)}(x')$ for some $j\in[l], j\neq i$. 
Hence, the fitness function for functionality testing could be 
\begin{equation}\label{eq:forward}
    \max_{j\neq i} (f^{(j)}(x') - f^{(i)}(x')) = \max_{j\neq i} g_{ji}(x'),
\end{equation}
i.e., any transformed input $x'$ with the greatest value of $\max_{j\neq i} g_{ji}(x')$ is considered the most promising one to expose the bug. Intuitively, the transformed input with a higher margin difference between other classes and the correct label is considered better and has a greater chance to alter the prediction. We refer to~\cref{eq:forward} as \emph{forward margin score}.

\paragraph{Loss score}
When the network is trained, a loss is used to measure the distance between the logit score $f(x)$ and the ground truth $i$. The training process minimizes the loss to fit the data to the label, i.e., if this measure is small, the $x$ is likely to be classified correctly. This naturally provides a measure for the bug: if the loss is big, then $x$ is likely to be misclassified. 
Based on this, we quantify how a muted input $x'$ is likely to cause the misclassification as follows,
\begin{equation}\label{eq:cross-entropy}
    CE(f(x'), i) = -\log\frac{e^{f^{(i)}(x')}}{\sum_{j=1}^n e^{f^{(j)}(x')}}.
\end{equation}
where we use the cross-entropy function, the most common and conventional choice of loss in classification tasks~\citep[chapter 3]{deep-learning}.
We refer to this score as \emph{forward loss score}. 

\subsection{Backward Fitness}\label{sec:diff}
To collect the logit score of input $x'$, one has to execute the DNN on $x'$. This can be expensive, especially when there are many input transformations, and the model is large. 
However, one distinguishable characteristic of neural networks compared to traditional software is that neural networks are differentiable. Recall that~\cref{eq:linear} defines a linear approximation of a function via gradients. This allows us to quickly estimate the values of the functions without executing the neural networks.

More formally, suppose we have an input $x$ and the transformed input $x'$, we can define $\Delta x = x'-x$. From~\cref{eq:linear}, we know that $f(x')\approx f(x)+(\Delta x)^TDf(x)$.
Based on this analysis, we introduce gradient-based surrogate versions of the two fitness scores proposed in~\cref{sec:forward}, \ie margin score and loss score.

\paragraph{Surrogate margin score}
For the margin score in~\cref{eq:forward}, the quantity we want to estimate is $g_{ji}(x')$. We can compute the the value of $g_{ji}$ at $x$, $g_{ji}(x)$, and the gradient at $x$, $Dg_{ji}(x)$, then we have
\begin{equation*}
\bar{g}_{ji}(x') = g_{ji}(x) + (\Delta x)^TDg_{ji}(x).
\end{equation*}
$\bar{g}_{ji}(x')$ is a linear approximation of the margin score ${g}_{ji}(x')$. Now, instead of identifying the input $x'$ that has the greatest value of $\max_{j\neq i} (f^{(j)}(x') - f^{(i)}(x'))$ ~(\cref{eq:forward}), we choose the input $x'$ that has the greatest value of
\begin{equation}\label{eq:backward-margin}
    \max_{j\neq i} (\bar{g}_{ji}(x')).
\end{equation}
We refer to this score as the \emph{backward margin score}, because we need the gradient to compute this approximation, and gradient computation requires backpropagation over the computation graph.

\paragraph{Surrogate loss score}
To approximate~\cref{eq:cross-entropy} at $x'$, we have
\begin{equation}\label{eq:ce-backward}
  \overline{CE}(f(x'), i) =  CE(f(x), i) + (\Delta x)^T\cdot D (CE)(f(x), i),  
\end{equation}
where $D (CE)(f(x), i)$ is the gradient of the cross-entropy function at $x$. We refer to this score as the \emph{backward loss score}. 


\paragraph{Computational efficiency} Using the linear approximation surrogate is particularly beneficial  
when the number of transformations is huge, and the number of model parameters is much larger than the dimensions of the inputs, which is always true for modern network architecture.
We compute the gradient with the original input once, and then for every transformation, we only need to compute the difference in the input domain, which is a much cheaper operation than executing the neural network with every transformed input.
In the supplementary material, we provide a running time analysis of gradient methods.

\paragraph{Analytical solution to linear approximation} The backward gradient surrogate can be more valuable when the transformation set is continuous and has a simple nice geometry. Because the set is continuous, theoretically, there are infinitely many transformations in the set. Therefore, forward-executing all transformed inputs becomes infeasible. However, if we use the gradient to approximate the function value, we can sometimes derive the optimal transformation among all possible ones to maximize the linear approximation in~\cref{eq:linear} analytically. 
For instance, if the transformation set is $\ell_2$ or $\ell_\infty$-transformations, the geometry of the sets is either Euclidean balls or hypercubes. To maximize the gradient inner product, one only needs to project the gradient to the $\ell_p$-balls. 
Specifically, if the perturbation set is an $\ell_2$-ball with radius $\epsilon$, the projection of the gradient inside the ball is
\begin{equation}\label{eq:l2}
    \frac{h}{\norm{h}_2}*\epsilon
\end{equation}
where $h$ is the gradient. 
If the perturbation set is an $\ell_\infty$-ball with radius $\epsilon$, then the projection of $h$ becomes
\begin{equation}\label{eq:linf}
\epsilon * \sign(h).
\end{equation}
\begin{proposition}\label{prop:opt}
    \Cref{eq:l2} is the optimal solution to the linear approximation in the $\ell_2$-ball; and \Cref{eq:linf} is the optimal solution to the linear approximation in the $\ell_\infty$-ball.
\end{proposition}
The proof is included in the supplementary materials.
This is the core idea of PGD, the state-of-the-art $\ell_p$-attacks, where the gradient comes from the loss-based score, similar to~\cref{eq:ce-backward}. Thus, when using gradient methods and the underlying transformation includes all small $\ell_p$-perturbations, we also use the gradient to generate the analytical solutions as in~\cref{eq:l2,eq:linf} to test DNN. 
\begin{remark}\label{rm:pgd}
In practice, PGD is a parameterizable multi-step adaptive projection of the gradients. On $\ell_\infty$-transformations, the one-step gradient projection coincides with the Fast Gradient Sign Method (FGSM) from the literature of adversarial examples~\citep{goodfellow2015explaining}.
\end{remark}

\subsection{Mixed Fitness}\label{sec:mixed}
Recall that in~\cref{rm:non-linear}, if $\Delta x$ is big, the approximation quality from gradients is poor. Therefore, we can also define a $\Delta x$-dependent metric such that, when $\Delta x$ is big, we use the forward score in~\cref{sec:forward}; and when $\Delta x$ is small, we consider the linear approximation in~\ref{sec:diff}. Accordingly, regarding the margin score, we define
\begin{equation}\label{eq:mm}
    MM(x') = \left\{ \begin{array}{ll}
           \max_{j\neq i} (g_{ji}(x')), & \mbox{$\norm{\Delta x}_p> \epsilon$}\\
            \max_{j\neq i} (\bar{g}_{ji}(x')), & \mbox{$\norm{\Delta x}_p \leq \epsilon$}\end{array} \right. 
\end{equation}
which is named \emph{mixed margin score}. Similarly, for the loss score, we have
\begin{equation}\label{eq:ml}
    ML(x') = \left\{ \begin{array}{ll}
           CE(f(x'), i), & \mbox{$\norm{\Delta x}_p> \epsilon$}\\
           \overline{CE}(f(x'), i), & \mbox{$\norm{\Delta x}_p \leq \epsilon$}\end{array} \right.
\end{equation}
which is named \emph{mixed loss score}. Note that the condition decision in~\cref{eq:mm,eq:ml} is a cheap operation. We only need to compute $\Delta x = x'-x$ and find the $\ell_p$-norm of all $\Delta x$. If we only consider small $\ell_p$ transformations, mixed scores correspond to their gradient-score counterparts. On the contrary, if all the transformation is large, mixed scores coincide with their forward-fitness counterparts. The rationale is that gradients measure the local variation of a function around a concrete input, and provide imprecise information when the input is allowed to change significantly. In this case, we should execute the DNN on the transformed input.

\paragraph{Theoretical takeaway}In fact, executing the DNN provides the most precise result, and should always be considered as a baseline. In an ideal scenario with unlimited computational resources, there would be no need to rely on gradient-based linear approximations: one can simply execute all transformed inputs and identify the buggy ones. Gradient-based linear approximation, while not as precise, presents a reasonable compromise when dealing with resource limitations and the underlying input change $\Delta x$ is small from~\cref{rm:non-linear}. PGD is the natural consequence of this approximation. However, one needs to understand the nature of input transformations to which gradient methods are applicable. Otherwise, the linear approximation can produce quite imprecise results.



%% file: eval.tex
\section{evaluation}

In this section, we evaluate the main claims that we made in~\cref{sec:thesis}:
\begin{enumerate}[parsep=1pt,itemsep=1pt,leftmargin=13pt,label={\arabic*.}]
    \item The directed metrics are effective in exposing bugs in functionality testing;
    \item Each metric is effective within its application scope and can have performance degradation outside the scope.
\end{enumerate} 
In addition, we conduct a comprehensive comparison between our directed metrics and diversity metrics, particularly when considering the small distortions as the input transformation, a critical aspect that has been overlooked by existing diversity-based DNN testing works~\citep{ATS,diversity-boosting}

For most testing frameworks, the ordinal relations of mutated inputs are more important than the absolute score values. Therefore, when comparing different metrics, our primary interest lies in assessing how closely these metrics align in ranking inputs. In particular, given that our forward fitness score is derived from logit scores, evaluating the similarity between other metrics and the forward fitness score enables us to understand the correlation between other metrics and changes in logit scores. This ranking similarity serves as the foundation of our work.

We then use various metrics to test the DNNs. We consider two testing scenarios, depending on whether ground-truth labels are available. When the ground-truth labels are available, we refer to the testing scenario as \emph{\true}. In \true, the goal is to generate inputs that cause misclassification as in the adversarial robustness literature~\citep{goodfellow2015explaining,pgd}. When ground-truth labels are unavailable, we refer to the testing scenario as \emph{metamorphic testing}. In this case, the testing rationale is that even if the ground-truth label is unavailable, if two inputs are metamorphic and classified differently, then at least one of the two inputs is misclassified. This testing scenario is considered in~\citet{metamorphic,diversity-boosting}. While metamorphic testing is not standard, it still falls into the category of functionality testing because the metamorphic inputs inducing divergent classifications also trigger misclassifications. 

Because the application scope is defined by the underlying input transformation, we evaluate the performances of the metrics with different types of transformations separately in all experiments.

More specifically, we want to answer the following research questions:
\begin{rqbox}{correlation}
    How closely do the metrics align in their ranking of the inputs? 
\end{rqbox}
\begin{rqbox}{natural}
    Are the directed metrics induced from our framework more effective than the diversity metric for exposing bugs in DNNs in standard testing tasks?
\end{rqbox}
\begin{rqbox}{adv}
    Are the directed metrics induced from our framework more effective than the diversity metric for exposing bugs in DNNs in metamorphic testing tasks?
\end{rqbox}



\subsection{Datasets and models}

We use four standard datasets. (1) MNIST~\citep{mnist} contains $10$ classes of grayscale handwritten images with $60,000$ training inputs and $10,000$ inputs. (2) SVHN~\citep{svhn} contains $10$ classes representing digits, $73257$ inputs for training, $26032$ digits for testing. (3) CIFAR10~\citep{cifar10} contains $60,000$ $32\times 32$ colour images in $10$ classes. There are $50000$ training images and $10000$ testing images. (4) CIFAR100~\citep{cifar10} is similar to CIFAR10, but has $100$ classes containing $600$ images each. There are $500$ training images and $100$ testing images per class. Among the four datasets, MNIST is grayscale inputs (1 channel) and the rest are RGB colorful inputs (3 channels).
We use three classical architectures from the computer vision community for our experiments: LeNet-1 and LeNet-5~\citep{lenet}, VGG-16~\citep{vgg}, and ResNet-9 and ResNet-18~\citep{resnet}. 
More details on datasets and models are summarized in~\cref{tab:setup} in the supplementary materials. 

\paragraph{Input Transformations}
We consider two sets of transformations on the input: (1) \textbf{small distortions (SDs)} that are specified by $\ell_\infty$-balls or $\ell_2$-balls with \emph{small} radii $\epsilon_\infty$ and $\epsilon_2$;
(2) \textbf{natural transformations} such as zooming and blurring~\citep{ATS}. 

Notice that natural transformations and SDs are not mutually exclusive. The difference is that for SDs, we would allow all possible small additive $\ell_p$ distortions, and since $\ell_p$-balls are continuous spaces, there are infinitely many transformations in theory. In practice, even if we discretize the space, the number of transformations is exponentially large with respect to the input dimension. As a result, if all small transformations are allowed, (1) for non-gradient methods, we can only sample the transformations within the $\ell_p$-balls; 
(2) for gradient methods, we can project the gradient to the $\ell_p$-balls directly as described in~\cref{eq:l2,eq:linf}.

We also report the average $\ell_2$-norm of $\Delta x$ for all the transformations on the test dataset. We choose $\ell_2$-norm instead of $\ell_\infty$-norm because the input space and inner product used in linear approximation~\cref{eq:linear} are naturally equipped with the $\ell_2$-norm. Also, the $\ell_2$-norm result can reflect the difference in $\ell_\infty$-norm.

\paragraph{Transformation specifications}
Since we test DNNs adaptively for $5$ iterations, we choose the parameters for input transformations so that all the transformations evaluated in recent works are attainable within $5$ iterations~\citep{gloro,cert_box,ATS}.

We choose $\epsilon_2 = 0.03$ and $\epsilon_\infty = 0.006$ on CIFAR10, SVHN and CIFAR100; and $\epsilon_2 = 0.32$ and $\epsilon_\infty = 0.06$ on MNIST. Because CIFAR10 and MNIST are standard datasets for adversarial learning, we use the radii from~\citet{cert_box,gloro}. More specifically, in~\citet{cert_box,gloro}, $\epsilon_2 = 1.58$ and $\epsilon_\infty = 0.3$ for MNIST and $\epsilon_2 \approx 0.14$ and $\epsilon_\infty \approx 0.03$ for CIFAR10. We divide them by $5$ to obtain the radii used in our evaluations. We choose the radii for CIFAR100 and SVHN the same as CIFAR10 because their input specifications are similar: $32\times 32$ RGB images. 


We consider $7$ types of natural transformations: shift, zoom, brightness, rotation, shearing, blur, and contrast ratio as used in~\citet{ATS}. Again, parameters are selected to ensure that all thes transformations are attainable within $5$ iterations.

\subsection{Experimental Design}
\paragraph{RQ1} We fix randomly selected transformations for each input in the test set, and 
then rank the transformations based on various metrics. We consider three types of transformations separately: (i) one with only SDs, (ii) one with only natural transformations, and (iii) one with both SDs and natural transformations.

For SDs, we randomly select $14$ $\ell_\infty$-perturbations and $14$ $\ell_2$-perturbations. For natural transformations, we generate $4$ transformations for each of the seven natural transformations as in~\citet{ATS}. For mixed transformations, we combine all $28$ SDs and all $28$ natural transformations together as the transformation candidates.

For every set of transformations, each fitness metric produces a ranked list of transformations, where the top ones are assumed to be more likely to expose bugs in the DNN. We then compute the rank similarities between the lists. For example, given a fixed set of inputs $(x_1, x_2, \ldots, x_n)$ generated from one specific testing datum $x$, each metric prioritizes them to generate a ranked list. We compare the two ranked lists to study the correlation between pairs of metrics. We use the Rank Biased Overlap (RBO) to measure this similarity between a pair of ranked lists. We choose RBO as the measurement because it is well-established~\citep{rbo}, and to the best of our knowledge, we are unaware of more suitable alternatives. The RBO score between two lists is the weighted average of overlaps between all top sublists from both lists, in which the weight takes the rank into account. The higher the score is, the more similar the two ranked lists are. One can then use RBO to measure how similar the two ranked lists are to study the correlation of the metrics. 

For each input $x$ from the testing dataset, we apply the input transformations $\Delta x_j$ to get a set of transformed inputs $\{x_j\}$ with $x_j= x + \Delta x_j$. We then use a pair of metrics to rank $\{x_j\}$ and compute the RBO score of the two ranked lists. As a result, if we fix a pair of metrics, for each datum $x_j$, we obtain an RBO score $r_j$. Suppose there are $n$ data in the test dataset (for example, $n=26032$ for SVHN). We then have $n$ RBO scores for each pair of metrics. 

\paragraph{RQ2} We evaluate the metrics in the adaptive testing scenario, i.e., a repeated loop involving the mutation of inputs, the evaluation of their fitness~\citep{tofu}, and the reception of feedback from the execution of inputs.
Let us fix a metric, a DNN $f$, and a seed input $(x^{(0)}, y)$ from the test dataset, where $0$ indicates this is the beginning of the first iteration and $y$ is the label of $x^{(0)}$. We consider all transformations $\Delta x^{(0)}_j$ over $x^{(0)}$, so we have a set of transformed inputs $\{x^{(0)}_j\}$ generated from $x^{(0)}+\Delta x^{(0)}_j$. We then rank the transformed inputs $\{x^{(0)}_j\}$ according to the metric. Suppose $x^{(0)}_0 = x^{(0)}+\Delta x^{(0)}_0$ is the best-transformed input according to the metric, we then retain $x^{(0)}_0$  as the seed input for the next iteration, i.e., $x^{(1)} = x^{(0)}_0$. 
In other words, we greedily choose the best transformation according to the metric for each input, and repeat this process for up to $5$ iterations. In each iteration, we collect $n$ best transformed inputs according to the metric, and use them as the seed inputs for the next iteration.

Recall that we designed metrics to detect DNN functionality bugs: inputs generated from semantics-preserving transformations and incorrectly classified. To evaluate the effectiveness of a metric, we run the DNN with the best-transformed inputs selected at the end of each iteration according to the metric, and record whether the transformed input is classified correctly with respect to the label $y$. Therefore, at the end of each iteration, we will get $\{c_1, \ldots, c_n\}$ with $c_i\in \{0, 1\}$. $c_i = 1$ means this input is correctly classified, and $c_i=0$ means this input is incorrectly classified. A metric that induces more incorrectly classified inputs has better bug-exposure capability.


In RQ2's experiment, we will consider three separate input transformations: natural transformations only, SDs only, and both natural transformations and SDs. The natural transformation-only scenario is similar to the evaluation in recent testing works~\citep{ATS,diversity-boosting}. SDs only can be viewed as the adversarial attack setting~\citep{pgd,gloro}. In particular, in the SDs-only scenario, adversarial examples created by existing attacks~\citep{pgd,goodfellow2015explaining} can be viewed as applications of our backward fitness scores. Because the goal of the experiment is to evaluate the bug-finding capability of different fitness, for non-gradient methods over SDs, their capability of bug-detection heavily relies on the number of sampled transformations. We, therefore, sample $140$ SDs instead of $28$ set in the evaluation of RQ1.

\paragraph{RQ3} We conduct similar experiments as those in RQ2, except that now the ground-truth labels are not available. The rationale for metamorphic testing is that metamorphic relations preserve the equivalence between inputs. Even if the ground-truth labels are unknown, if two metamorphic inputs are classified differently, at least one of the inputs is classified incorrectly, so this exposes a bug. As a result, the testing goal is to apply metamorphic relations on input and change the prediction.

We can easily adapt our metrics in the metamorphic testing scenario. Because the goal is to alter the prediction for the original inputs rather than ground-truth labels, therefore, instead of using correct labels, we use pseudo-labels predicted by the model on the original input, even though they are not necessarily correct. The goal for this testing setting is to prioritize transformed inputs that are more likely to alter the pseudo-label predictions.

The experiments are similar to RQ2's adaptive testing case. However, instead of measuring whether the input is correctly classified relative to the ground-truth label, we measure the accuracy relative to the pseudo-labels generated by the original input (without any transformations). This measurement aligns with the metamorphic testing goal, i.e., to generate metamorphic inputs that induce different predictions. We refer to the aggregation of this measurement as pseudo-accuracy.


\paragraph{Statistical analysis} 
For all the aggregated results, we compute their average at $95\%$ confidence intervals. The statistical analysis details can be found in~\cref{sec:stat} in the supplementary material.

\subsection{Experimental Specifications}\label{sec:exp-spec}
\paragraph{Baseline} We use three recent diversity-based testing works as baselines:
\begin{enumerate}[parsep=1pt,itemsep=1pt,leftmargin=13pt,label={\arabic*.}]
    \item Boosting Diversity:~\citet{diversity-boosting} is our major baseline. It is built upon the assumption that if the divergence between two executions is significant, then the inputs might be classified differently. 
    Transformed inputs with a great difference should be prioritized.
    As noted in~\citet{diversity-boosting}, as one uses features extracted at the deeper layer of DNNs, the prioritization accuracy increases, but the execution time increases. 
    To find the balance between these two, the standard practice is to choose the layer in the middle of the target DNN.
    We follow this guideline but also include a variant as a stronger baseline; we use the logit layer as the representation layer, which provides better prioritization estimation than the mid-intermediate layer.
    \item NC:~\citet{deep_gauge} monitor and gauge the intermediate neuron activities covered by current executions at various granularity levels. They profile the range of neuron outputs into $k$-multisections of the major function region and corner-cases regions during the training, then identify the ratio of sections covered under test-time executions.
    We follow the modified NC measure in~\citet{diversity-boosting} as one of our baselines. 
    \item ATS:~\citet{ATS} is a recently proposed metric for input selection, i.e., choosing a small yet diverse subset of inputs from a large set. 
    We adapt it to our scenario by feeding an input and its transformed mutants to the selection algorithm one at a time and then rank the inputs based on their metric. We include this variant as another baseline.
\end{enumerate}

We use the following abbreviations for the methods considered in the evaluation: 
\textbf{FM} stands for forward margin score~(\cref{eq:forward});
\textbf{FL} stands for forward loss score~(\cref{eq:cross-entropy}); \textbf{BM} stands for backward margin score~(\cref{eq:backward-margin}); \textbf{BL} stands for backward loss score~(\cref{eq:ce-backward}); \textbf{MM} stands for mixed margin score~(\cref{eq:mm}); \textbf{ML} stands for mixed loss score~(\cref{eq:ml}); \textbf{BD} stands for the standard tool from~\citet{diversity-boosting}, \textbf{BD-F} stands for a variant that uses the logit layer to compute the divergence. \textbf{NC} stands for NC method using the layer in the middle of DNN architecture, \textbf{ATS} stands for the adapted tool from~\citet{ATS}.

To summarize, our metrics (FM, FL, BM, BL, MM, ML), ATS, and BD-F only utilize the logit layer, while BD and NC use the intermediate layer neuron activations. 
We choose the same intermediate layer for both BD and NC. 
We do not include evaluations of ATS on the CIFAR100 dataset since the fault pattern computation is cubic to the number of classes, which is too slow with 100 classes.

Additional experimental setup can be found in~\cref{sec:app-spec} in the supplementary material.

\section{Results and Discussion}

In this section, we present selected experimental results due to space limits and discussions on the results. Notice that even though we only showcase selected results in the main text, the observations presented hold for \emph{all} experiments that we conducted, and the complete results are in the supplementary material. 
\input{rq1_table}

\subsection{RQ1}

\paragraph{RBO baseline}A high RBO score indicates that two ranked lists are positively correlated and similar. We benchmark $3$ RBO scores: (i) two same-ranked lists (denoted as \emph{identical}); (ii) two randomly ranked lists (denoted as \emph{random}); (iii) two opposite-ranked lists (denoted as \emph{opposite}). Because the scores are list-length dependent, we use $10000$ CIFAR10 testing inputs to benchmark these scores on lists with lengths $28$ (corresponds to $\ell_p$ and natural transformations only) and $56$ (corresponds to mixed transformations). With these benchmarks, summarized in~\cref{tab:rbo-benchmark}, we can interpret the results more precisely.
\begin{table}[H]
\begin{center}
\vspace{-0.3cm}
\caption{RBO score benchmark}\label{tab:rbo-benchmark}
\vspace{-0.2cm}
\begin{adjustbox}{width=0.8\columnwidth,center}
\begin{tabular}{ c|ccc } 
\toprule
\bf Length & \bf Identical & \bf Random & \bf Opposite \\
\hline \hline
  28 & $1.0\pm0.0$  & $0.518\pm 0.001$ & $0.324\pm 0.0$ \\ 
  \cline{1-4}
  56 & $1.0\pm0.0$  & $0.509\pm 0.001$ & $0.316\pm 0.0$ \\ 
  \bottomrule
\end{tabular}
\end{adjustbox}
\vspace{-0.3cm}
\end{center}
\end{table}

As a result, if the RBO score is less than $0.519$ for lists of length $28$ or $0.510$ for lists of length $56$, the two ranks are negatively correlated. 

The RBO scores between different ranking methods are summarized in~\cref{tab:rq1-c} in the supplementary material, and selected results are presented in~\cref{tab:rq1}. 

\paragraph{Observations}
1. The forward loss and margin scores are always very positively correlated.

2. Diversity metrics and the loss score are quite weakly correlated when the input transformation induces small changes in the inputs.  This correlation is stronger when the underlying input transformations induce large changes.

3. Forward scores and backward scores are strongly positively correlated when the underlying transformation is only SDs; however, for natural transformations, the norms induced from $\Delta x$ are fairly large, and the correlations between forward and backward scores are much weaker.

\paragraph{Implications}
1. All the scores can be more effective in exposing functionality bugs than random prioritization. However, When applying SDs on the input, the correlation between diversity metrics and the loss change in the logit layer is quite weak. This raises concerns about their effectiveness in detecting the change in the logit layer that leads to a better chance of misclassification. 

2. Both margin and loss scores can quantify how a DNN makes predictions, so their rankings of inputs to be classified incorrectly are very similar. 

3. Gradient-based scores are excellent surrogates for forward scores only when the underlying transformations are SDs.


\input{rq2-fig}

\subsection{RQ2}
The complete results are in~\cref{tab:rq2-c1,tab:rq2-c2,tab:rq2-c3} in the supplementary material. For brevity, we discuss the results of the VGG model on CIFAR100, as the results of other models are similar. We plot the trend of accuracy in~\cref{fig:rq2-2}.

\paragraph{Observations}In the standard testing scenario, when the underlying transformations are different, the performances of the metrics vary significantly:
\begin{enumerate}[parsep=1pt,itemsep=1pt,leftmargin=13pt,label={\arabic*.}]
    \item When only SDs are presented, our backward margin (\textbf{BM}) and backward loss (\textbf{BL}) scores are by far the most effective metrics for identifying inputs that induce functionality bugs in DNNs. In particular, after the third iteration, \textbf{BM} and \textbf{BL} are close to 10 times more effective than the diversity metrics (i.e., \textbf{BD}, \textbf{BD-F}, and \textbf{NC}). In fact, diversity metrics are barely useful for finding bug-inducing inputs when only considering the SDs. Those that are highly ranked by diversity metrics rarely lead DNNs to misclassifications. Even our forward scores (i.e., \textbf{FM} and \textbf{FL}) are more effective than diversity metrics. Again, we emphasize that not accounting for SDs as an input transformation is not a valid excuse for existing diversity-based testing methods since such data points can certainly arise in the real world as inputs to DNNs.     

    \item When we only consider natural transformations, the forward scores produce the best results. 
    Since the accuracy of BD(-F)'s is close to FM's and FL's, we provide a zoomed view of their averaged accuracy together with the 95\% confidence interval in~\cref{tab:rq2-select}. The statistical analysis shows that at the third iteration, the models' accuracy is around 10-20 times lower when handling inputs identified by FM or FL compared to inputs identified by BD(-F), and the margin becomes even wider at later iterations.    
    In the meantime, the gradient-based scores are weaker than diversity-based metrics, though inputs that are highly ranked by gradient-based scores also consistently decrease models' accuracy. 
    \item When both SDs and natural transformations are allowed, our mixed scores are the most effective metrics.
\end{enumerate}

\input{rq2_table}

\paragraph{Implications}
1. When only SDs are allowed, our backward scores are significantly more effective than diversity metrics for finding inputs that trigger bugs in DNNs. In the natural transformations-only scenario, our forward scores are the most effective in finding bug-inducing inputs. These two observations together establish the main thesis that DNN testing is a directed testing task, and the likelihood of an input inducing bugs in DNNs can be accurately measured by our directed metrics.




2. From RQ1's result, we can see that gradient surrogates are very positively correlated when input transformations are small distortions. Moreover, the linear approximation allows for the direct derivation of the optimal transformation when all SDs are allowed. This observation is consistent with our theoretical analysis and validates the application of gradient methods for SDs as in the adversarial attack case from DL literature~\citep{szegedy2014intriguing,pgd,goodfellow2015explaining}. 

3. Overall, for all input transformations, our metrics perform the best in finding inputs that cause models to misclassify. We also demonstrate the scope of each metric as presented in~\cref{sec:thesis}. As shown by our results, applying a metric outside of its scope results in a drastic degradation of its performance.


\subsection{RQ3}
\input{rq3_table}
The complete results are presented in~\cref{tab:mt-nat,tab:mt-sd,tab:mt-mix} in the supplementary material. We showcase the MNIST LeNet-5 model (the results for other models are similar) and plot the pseudo-accuracy changes in~\cref{fig:rq3-1}.
\input{rq3-fig}

\paragraph{Observations}The overall performances of the metrics are similar to the standard testing scenario: e.g., when only SDs are allowed, the diversity metrics are rather ineffective for finding bug-inducing inputs; in the natural transformations-only case, the forward metrics give the best results. 
We also provide a zoomed view of natural transformations-only experiment in~\cref{tab:rq3-select} with the 95\% confidence intervals. We still observe that our forward metrics are more effective than the BD variants for finding inputs that trigger bugs in DNNs.

\paragraph{Implications} Overall, the performances of our metrics are similar to the standard testing. This implies that our directed metrics are also effective when we approach DNN functionality testing as metamorphic testing. In essence, we use a pseudo-label predicted by the model as the ground truth, and the results show that our metrics are more effective than diversity metrics in finding inputs that alter the models' prediction of the original inputs. Again, this demonstrates that the metamorphic testing of DNNs should also be treated as a directed rather than a diversity testing task. 

\subsection{Empirical highlights}
Our evaluation reveals the following key points:
\begin{enumerate}[parsep=1pt,itemsep=1pt,leftmargin=13pt,label={\arabic*.}]
    \item Diversity metrics are significantly ineffective for finding bug-inducing inputs when the input transformation is restricted to small distortions. This testing scenario has been overlooked by existing diversity-based DNN testing methods~\citep{diversity-boosting,ATS}.
    \item Though both SDs and natural image transformations preserve the visual semantics, they have distinct mathematical properties and should be addressed separately for testing purposes.  Applying metrics outside of their scope would result in a significant degradation in their effectiveness. 
    \item Regardless of the type of the underlying input transformations, our directed testing metrics are more effective in finding inputs that trigger the functionality bugs in DNNs than diversity metrics. Therefore, we substantiate the main thesis of this paper: DNN testing should be treated as a directed test task.
\end{enumerate}

%% file: rq1_table.tex
\begin{table*}[ht]
\centering
\caption{Similarity measures (RBO) between ranked lists according to various prioritization metrics. The first column specifies the model, dataset and accuracy on the test dataset. The second column specifies what transformations are used for the experiment. The third column norm measures the average $\ell_2$-norm of all transformations. The rest of the columns measure the similarity between pairs of metric scores. \colorbox{blue!10}{We omit the confidence interval when it is $<0.005$}.}
\begin{adjustbox}{width=1\textwidth,center}
\begin{tabular}{l|l|c|c|c|c|c|c|c|c}
\toprule
\bf{Data, Model,} & \bf{Transfor-} & \multirow{2}{*}{\bf Norm} & \multirow{2}{*}{\bf FM\&FL} & \multirow{2}{*}{\bf FM\&BM} & \multirow{2}{*}{\bf FL\&BL} & \multirow{2}{*}{\bf FL\&BD} & \multirow{2}{*}{\bf FL\&BD-F} & \multirow{2}{*}{\bf FL\&ATS} & \multirow{2}{*}{\bf FL\&NC} \\ 
\bf Accuracy & \bf mation & & & & & & & & \\ \hline \hline

SVHN & SD & 0.11  & 0.90 & 0.95 & 0.93  & 0.54 & 0.55 & 0.58   & 0.54 \\
\cline{2-10}
ResNet-9 & Natural & $8.60 \pm 0.02$  & 0.95  & 0.57 &  0.55 & 0.71  & 0.78  & 0.76  &  0.72  \\ \cline{2-10}
95.90\% & Mixed  &  $4.22\pm 0.01$  & 0.94  & 0.68  & 0.66  & 0.69  & 0.72  & 0.72  & 0.70  \\ \cline{1-10}

CIFAR10, & SD & 0.11 & 0.91& 0.93 & 0.92 & 0.56 & 0.58 & 0.56 & 0.56 \\\cline{2-10}
VGG, & Natural & $11.42\pm 0.04$ & 0.96& 0.57 & 0.55 & 0.75 & 0.79 & 0.71  & 0.73 \\
\cline{2-10}
91.48\% & Mixed  & $5.67\pm 0.02$ & 0.95 & 0.68 & 0.66 & 0.71 & 0.73 & 0.69 & 0.70\\
\bottomrule
\end{tabular}
\end{adjustbox}
\vspace{0.5em}
\label{tab:rq1}
\vspace{-0.5em}
\end{table*}

%% file: rq2-fig.tex
\begin{figure*}[t]
     \centering
     \begin{subfigure}[b]{0.32\textwidth}
         \centering
         \includegraphics[width=\textwidth]{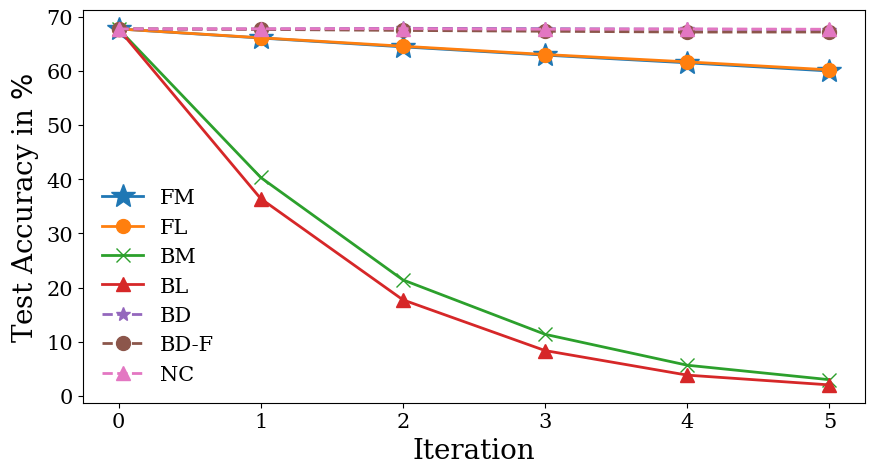}
         \caption{SDs only}\label{fig:cifar100-sd}
     \end{subfigure}
     \hfill
     \begin{subfigure}[b]{0.32\textwidth}
         \centering
         \includegraphics[width=\textwidth]{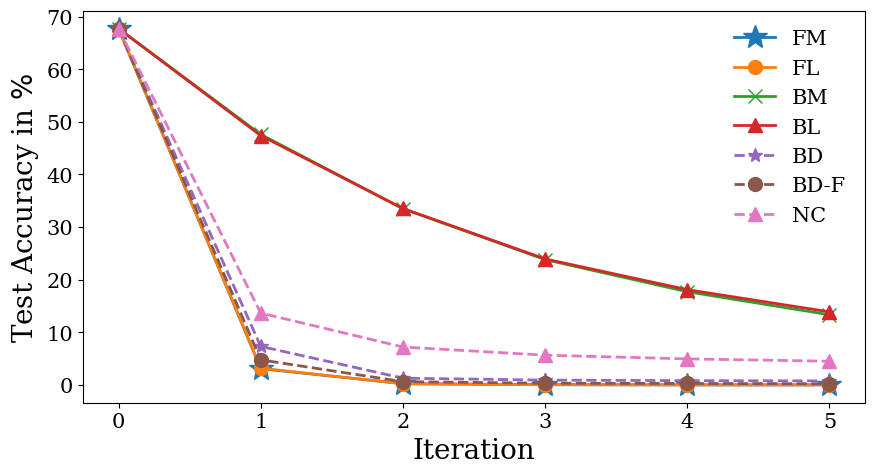}
         \caption{Natural transformations only}\label{fig:cifar100-nat}
     \end{subfigure}
     \hfill
     \begin{subfigure}[b]{0.32\textwidth}
         \centering
         \includegraphics[width=\textwidth]{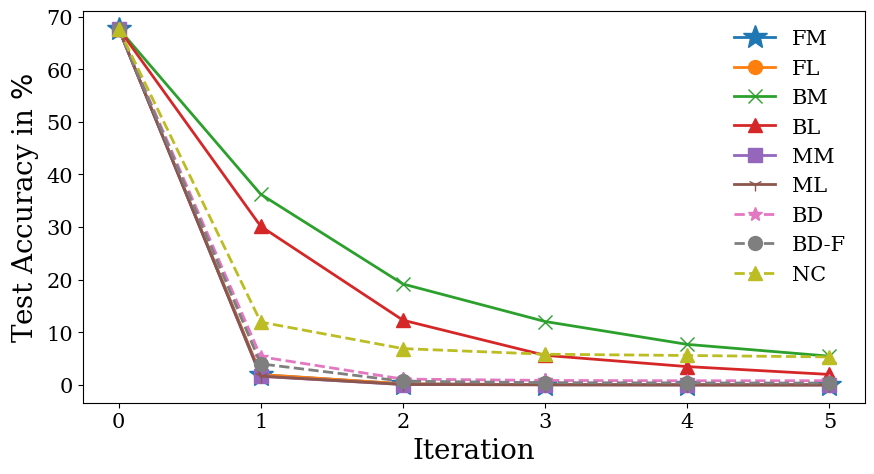}
         \caption{SDs and natural transformations}\label{fig:cifar100-mix}
     \end{subfigure}
        \caption{Accuracy change of CIFAR100 VGG in adaptive standard testing using various metrics, with different input transformations: only natural transformations, only small distortions, and both natural transformations and small distortions.}
        \label{fig:rq2-2}
\end{figure*}

%% file: rq2_table.tex
\begin{table*}
\begin{center}
\caption{Adaptive standard testing of each metric: the first column specifies the data, model, accuracy and input transformations. The second column specifies which metric is used for testing. The rest of the columns are the accuracy of tests prioritized by each metric during each iteration in adaptive testing. We also include the 95\% confidence interval of the averaged accuracy.}\label{tab:rq2-select}
\begin{adjustbox}{width=\textwidth,center}
\begin{tabular}{c|c|ccccc}
\toprule
\bf Data, Model, Accuracy, Transformations & \bf Fitness & \bf Iter 1 &\bf Iter 2 & \bf Iter 3  & \bf Iter 4 & \bf Iter 5 \\
\hline \hline
\hfil CIFAR100& \hfil FM & \hfil $3.10\pm 0.34$\% & \hfil $0.24\pm 0.10$\% &  \hfil $0.05\pm 0.04$\% & \hfil $0.01\pm 0.01$\% & \hfil $0.01\pm 0.01$\%  \\
\cline{2-7}
\hfil VGG& \hfil FL  &  \hfil $3.18\pm 0.34$\% & \hfil $0.23\pm 0.09$\% & \hfil $0.04 \pm 0.04$\% & \hfil $0.02\pm 0.02$\% & \hfil $0.01\pm 0.01$\%  \\
\cline{2-7}
\centering 67.73\%, & \hfil BD & \hfil $7.33\pm 0.51$\% & \hfil $1.29\pm 0.22$\% & \hfil $0.92\pm 0.19$\% & \hfil $0.82\pm 0.18$\% & \hfil $0.77\pm 0.17$\% \\
\cline{2-7}
\hfil Natural& \hfil BD-F & \hfil $4.77\pm 0.42$\% & \hfil $0.63\pm 0.16$\% & \hfil $0.33\pm 0.11$\% & \hfil $0.28\pm 0.10$\% & \hfil $0.23\pm 0.09$\% \\
\bottomrule
\end{tabular}
\end{adjustbox}
\end{center}
\end{table*}

%% file: rq3_table.tex
\begin{table*}
\begin{center}
\caption{Adaptive metamorphic testing of each metric: the first column specifies the data, model, pseudo-accuracy and input transformations. The second column specifies which metric is used for testing. The rest of the columns are the pseudo-accuracy of tests prioritized by each metric during each iteration in adaptive testing. We also include the 95\% confidence interval of the averaged pseudo-accuracy.}\label{tab:rq3-select}
\begin{adjustbox}{width=\textwidth,center}
\begin{tabular}{c|c|ccccc}
\toprule
\bf Data, Model, Accuracy, Transformations & \bf Fitness & \bf Iter 1 &\bf Iter 2 & \bf Iter 3  & \bf Iter 4 & \bf Iter 5 \\
\hline \hline
\hfil MNIST& \hfil FM & \hfil $8.65\pm 0.55$\%& \hfil $0.10\pm 0.06$\%& \hfil $0.01\pm 0.01$\%& \hfil $0.00\pm 0.00$\% & \hfil $0.00\pm 0.00$\% \\
\cline{2-7}
\hfil LeNet-5& \hfil FL  & \hfil $8.72\pm 0.55$\%& \hfil $0.12\pm 0.07$\%& \hfil $0.01\pm 0.01$\%& \hfil $0.00\pm 0.00$\% & \hfil $0.00\pm 0.00$\% \\
\cline{2-7}
\centering 100.0\%, & \hfil BD & \hfil $29.37\pm 0.89$\%& \hfil $12.71\pm 0.65$\% & \hfil $9.81\pm 0.58$\% & \hfil $9.06\pm 0.56$\% & \hfil $8.43\pm 0.54$\%\\
\cline{2-7}
\hfil Natural& \hfil BD-F & \hfil $18.53\pm 0.76$\%& \hfil $5.01\pm 0.43$\% & \hfil $2.39\pm 0.30$\% & \hfil $1.90\pm 0.27$\% & \hfil $1.57\pm 0.24$\%\\
\bottomrule
\end{tabular}
\end{adjustbox}
\end{center}
\end{table*}

%% file: rq3-fig.tex
\begin{figure*}[t]
     \centering
      \begin{subfigure}[b]{0.32\textwidth}
         \centering
         \includegraphics[width=\textwidth]{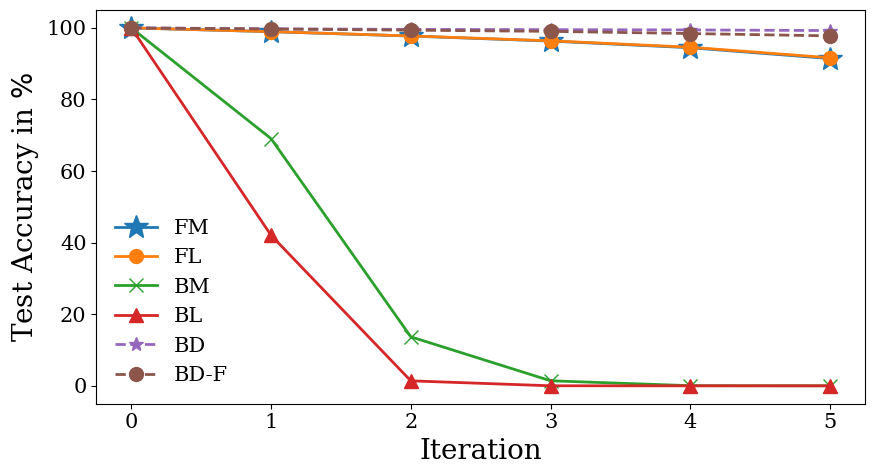}
         \caption{SDs only}
     \end{subfigure}
     \hfill
     \begin{subfigure}[b]{0.32\textwidth}
         \centering
         \includegraphics[width=\textwidth]{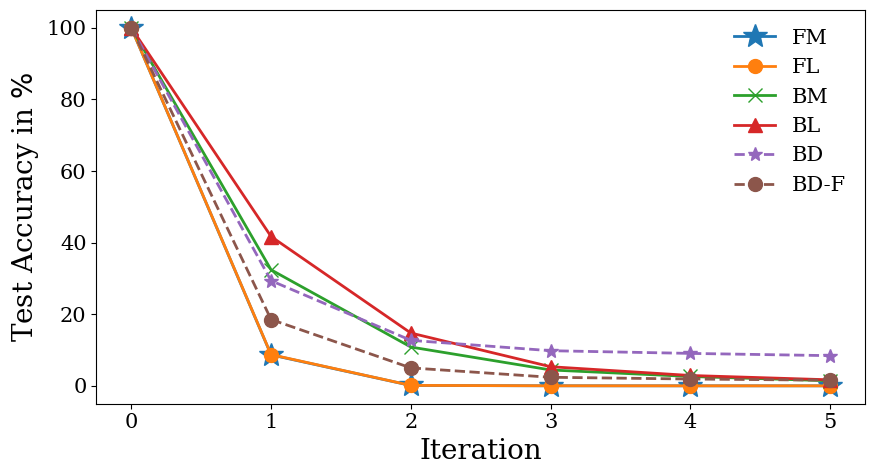}
         \caption{Natural transformations only}
     \end{subfigure}
     \hfill
     \begin{subfigure}[b]{0.32\textwidth}
         \centering
         \includegraphics[width=\textwidth]{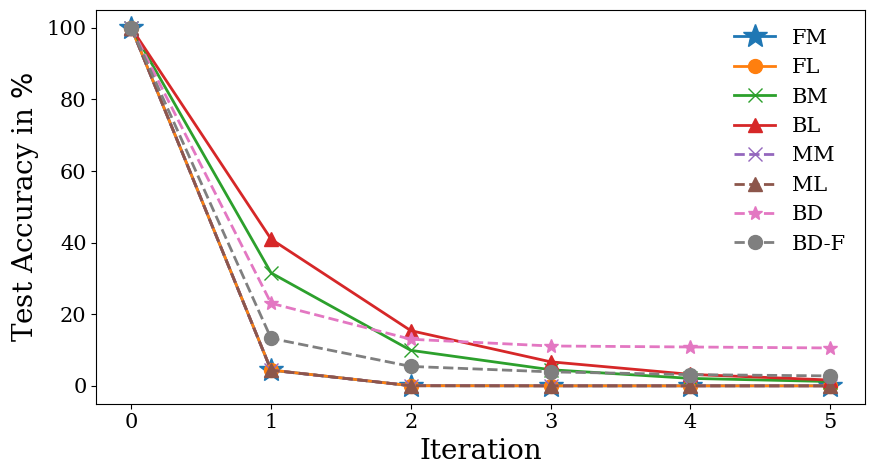}
         \caption{SDs and natural transformations}
     \end{subfigure}
        \caption{Pseudo-accuracy change of MNIST LeNet-5 in adaptive metamorphic testing using various metrics, with different input transformations: only natural transformations, only small distortions, and natural transformations and small distortions.}
        \label{fig:rq3-1}
\end{figure*}

%% file: conclusion.tex
\section{Threats to validity}
\paragraph{Internal validity}Each number presented is based on the statistical analysis of at least $10,000$ measurements. We justified most of the evaluation design choices except for iteration numbers. In our evaluation, we used $5$ iterations to study the adaptive testing scenario. This number is not fully justified, and the parameters used in the evaluation, such as $\ell_p$ radii and natural input transformation parameters, depend on this iteration number. However, the $5$ iteration experiment showcases examples of adaptive testing, and in each iteration, our metrics are always the best and only BD-F is relatively close to our metrics for standard testing. We re-conduct the evaluation for standard testing with iterations $3$ and $8$ on our metrics and BD(-F) and summarize the result in the supplementary material. The conclusion is still the same: our metrics are superior to BD(-F) for different iteration numbers.

\paragraph{External validity}
Our empirical results are based on $4$ datasets and $3$ model architectures, with various numbers of inputs, image channels (RGB v.s. grayscale), classes, and sizes. This diversity of inputs and models enhances the generalizability of our evaluations.
We restrict the scope of our work to testing functionality bugs on DNNs and claim that these tasks are directed. This is because DNNs are specialized numerical programs and their mathematical semantics is simple. 
However, for sophisticated DL systems such as hybrid systems, which contain both DNN and traditional software, their semantics can be complicated, and testing them with designed diversity measures might be more appropriate.

\section{Related Work}
We claim that merely increasing diversity is not the ideal framework for DNN functionality testing. However, we do not claim that diversity is useless for DL testing. It depends on our testing goal. A natural diversity issue of DL systems is fairness and bias~\citep{fairness,nn-bias}: Between subgroups, DL systems contain performance discrepancies and make biased decisions towards a certain subgroup. If the tests were only from a subgroup, then potential fairness and bias issues would be concealed. Therefore, designing meaningful diversity measures can help expose and alleviate the bias and fairness issues in DL systems.

In this work, we restrict our scope to functionality bugs, i.e., inputs that are misclassified according to human perception. However, DL bugs can manifest in various forms. In~\citet{debug}, the authors categorize model bugs into three types: model contamination, data contamination, and test-time contamination. Model contamination is caused by errors in the model parameters, for example, incorrectly initialized parameters; data contamination comes from defects in the training data; and test-time contamination is caused by shifts in test data.  
Functionality bugs can be viewed as test-time contamination: our data are transformed at test time. However, there are many more types of bugs in the DL system to be addressed and software testing techniques can be applied. For example, model contamination is often caused by API misuses, which has been studied in software engineering~\citep{api-misuse1,api-misuse2}.


\section{Conclusion}In this paper, we uncover the shortcomings of diversity metrics in DNN functionality testing and advocate that it should be addressed using the directed testing framework because it is well-defined and specific. Moreover, we propose several concrete metrics to quantify the potential faultiness of inputs and carefully characterize their scopes. We find that metrics perform drastically differently across input transformations and call for careful scrutiny of these methods.


%% file: appendix.tex
\newpage
\appendix

%




\section{A linear approximation illustration}

\input{linear}

\section{Gradient methods running-time analysis}
Suppose there are $N$ transformations for an input, a forward execution takes $F$ time, a backward gradient execution takes $B$ time, which is usually similar to $F$, and the inner-product computation takes $I$ time. To compute the forward fitness score, we need $N$ forward executions, which take $N F$ time. To compute the backward fitness score, we need $1$ forward execution to compute $f(x)$ and $1$ backward execution to compute the gradient $Df(x)$, and then $N$ inner-product computations to compute $(\Delta x)^T D(f)(x)$, which is usually very cheap compared to DNN executions. Brought together, the backward fitness score takes $F+B+NI$ time to finish.

\section{Proof of proposition 3.1}
\begin{proof}
    Let us fix an input $a$. The goal is to maximize the linear approximation of a function $f(x)$ at $a$ within the $\ell_p$-balls with radii $\epsilon$ centered at $a$.
    
    Because the balls are centered at $a$, each point in the ball can be expressed as $a+\Delta x$ with $\norm{\Delta x}_p\leq \epsilon$, and the linear approximation at the point is:
    \[f(a) + (\Delta x)^T Df(a).\]
    Let us use $h$ to denote $Df(a)$. To maximize this term, we only need to maximize $(\Delta x)^T h$. 

    If $\norm{\Delta x}_2\leq \epsilon$, because $(\Delta x)^T h = \norm{\Delta x}_2 \norm{h}_2 \cos{\theta}$, where $\theta$ is the angle between the two vectors $\Delta x$ and $Df(a)$, we know that $\norm{\Delta x}_2\leq \epsilon$ and $\cos \theta\leq 1$, then $(\Delta x)^T Df(a)\leq \epsilon \norm{h}_2$. On the other hand, if $\Delta x = \frac{h}{\norm{h}_2}\epsilon$, then $(\Delta x)^T h = \frac{h^T h}{\norm{h}_2}\epsilon = \epsilon\norm{h}_2$. This shows that \Cref{eq:l2} is the optimal solution to the linear approximation in the $\ell_2$-ball. 
    
    Notice that in above proof, we only need $(\Delta x)^T h \leq \norm{\Delta x}_2 \norm{h}_2$, which can be directly derived from the Cauchy–Schwarz inequality. If we generalize the Cauchy–Schwarz inequality to Hölder's inequality, we can immediately get the result for the $\ell_\infty$-case.

    More formally, if $\norm{\Delta x}_\infty\leq \epsilon$, then by Hölder's inequality, $(\Delta x)^T h\leq \norm{\Delta x}_\infty \norm{h}_1 \leq \epsilon \norm{h}_1$. On the other hand, if $\Delta x = \epsilon * \sign(h)$, then $(\Delta x)^T h =  \epsilon * \sign(h) h = \epsilon \sum |h_i| = \epsilon \norm{h}_1$. This shows that \Cref{eq:linf} is the optimal solution to the linear approximation in the $\ell_\infty$-ball.
    
\end{proof}

\section{Statistical Data Analysis}\label{sec:stat}
In RQ1, for each single datum, we measure the RBO score, a continuous value between $[0, 1]$. For RQ2 and RQ3, for every datum, we measure the accuracy, a binary value from $\{0,1\}$, which stands for wrong and correct classification, respectively. As a result, for RQ1, we collect a set of RBO scores $\{r_1, \ldots, r_n\}$, where $n$ is the number of testing inputs and $r_i\in [0, 1]$; and for RQ2 and RQ3, in each iteration for each metric, we collect a set of accuracy values $\{c_1, \ldots, c_n\}$, where $c_i\in \{0,1\}$.

We conduct statistical data analysis on two types of random variables: continuous-valued and binary-valued. Because each statistic is from thousands of measurements, i.e., each datum in the testing dataset produces a measurement, we use the normal distribution to model the data distribution due to central limit theorem~\citep{stat} and the Scipy.stats~\citep{scipy} and statsmodels~\citep{statsmodels} packages to compute the confidence interval.

\section{Additional experimental specifications}\label{sec:app-spec}

\begin{table}[th]
\begin{center}
\caption{Datesets and Models}
\vspace{-0.2cm}
\begin{adjustbox}{width=1\columnwidth,center}
\begin{tabular}{ cc|ccc } 
\toprule
\bf Dataset & \bf Description & \bf Model & \bf Layer & \bf \# Parameters \\  
\hline \hline
 \multirow{2}{*}{MNIST} & 70000 $28\times 28$  & LeNet-1 & 5 & 3246\\ 
  & grayscale images & LeNet-5 & 7 & 61706\\
  \cline{1-5} 
  \multirow{2}{*}{SVHN} & 99289 $32\times 32$  & VGG-16 & 21 & 14990922\\ 
  & RGB images & ResNet-9 & 9 & 2274880\\
  \cline{1-5}
  \multirow{2}{*}{CIFAR10} & 60000 $32\times 32$  & VGG-16 & 21 & 14990922\\ 
  & RGB images & ResNet-18 & 18 & 11173962\\
  \cline{1-5} 
  \multirow{2}{*}{CIFAR100} & 60000 $32\times 32$  & VGG-16 & 21 & 15037092\\ 
  & RGB images & ResNet-18 & 18 & 11220132\\
  \bottomrule
\end{tabular}
\label{tab:setup}
\end{adjustbox}
\end{center}
\vspace{-0.3cm}
\end{table}

\paragraph{Server specification} All the experiments are run on a workstation with forty-eight Intel\textsuperscript{\textregistered} Xeon\textsuperscript{\textregistered} Silver 4214 CPUs running at 2.20GHz, and 258 GB of memory, and eight Nvidia GeForce RTX 2080 Ti GPUs. Each GPU has 4352 CUDA cores and 11 GB of GDDR6 memory.

\paragraph{Training specifications} We use batch size $500$, the SGD optimizer with learning rate $0.1$, momentum $0.9$, decay $5e^{-4}$ to train the networks, and each model is trained for $100$ epochs. 
To ensure reproducibility, we also explicitly specify seeds for all random functions.

\input{appendix_rq1_tab}

\input{appendix_rq2_tab}

\input{appendix_rq3_tab1}

\input{appendix_rq3_tab2}

\input{mt-tab}

\input{mt-tab-sd}

\input{mt-tab-mix}


\section{Different Iteration Numbers}
We rerun the experiments for RQ2 on our forward metrics and BD(-F), as if the iterations were 3 or 8. Therefore, we also adjust the parameters for the input transformations accordingly. For example, if for the 5 iterations, the shift step is 0.3, then for the 3 iteration experiment, we adjust the step to 0.5 and for the 8 iteration experiment, we adjust the step to 0.1875. The rationale is that we want to make sure the total step size for all experiments are the same, i.e., $0.3\times 5= 0.5\times 3 = 0.1875\times 8$. The results are presented in~\cref{tab:iter3,tab:it8}. Notice that the result is consistent with our 5-iteration experiments.
\input{tab-iter3}

\input{tab-iter8}

%% file: linear.tex
    \begin{figure}{ht}
    \vspace{-1em}
    \centering
        \begin{tikzpicture}
            \begin{axis}[
                axis lines=middle,
                y=1.2cm,
                x=1.2cm,
                xmax=2.2,
                xmin=-1.5,
                xtick={-1,0, 2},
                ymin=-0.8,
                ymax=2.2,
                ytick={0,1, 2},
                width=2cm
            ]

            \addplot [domain=-3:3, samples=100,
            thick, red] {0.5*x^2};

            \addplot [domain=-3:3, samples=100,
            blue] {0.5*x-0.125};

            \node at (axis cs:0.5,0) [circle, scale=0.3, draw=black!80,fill=black!80] {};

            \node at (axis cs:0.8,0) [circle, scale=0.3, draw=black!80,fill=black!80] {};

            \node at (axis cs:1.6,0) [circle, scale=0.3, draw=black!80,fill=black!80] {};

            \tikzstyle{dashed}=[dash pattern=on 3pt off 3pt,color=blue]
            \draw[dashed] (0.5,0.125) node[left] {} -- (0.5,0) node[below] {$a$};
            \draw[dashed] (0.8,0.32) node[below] {} -- (0.8,0) node[below] {$x_1$};
            \draw[dashed] (1.6,1.28) node[below] {} -- (1.6,0) node[below] {$x_2$};
        
        \end{axis}
        \end{tikzpicture}
        \caption{An illustration of linear approximation of $f(x) = 0.5*x^2$ (\textcolor{red}{red curve}) at $a = 0.5$. One can use the gradient of $f$ to construct the linear approximation as in the \textcolor{blue}{blue line}, which is close to $f$ when $x_1$ is close to $a$, but distant when $x_2$ is far from $a$.}\label{fig:linear_approx}
        
        \end{figure}
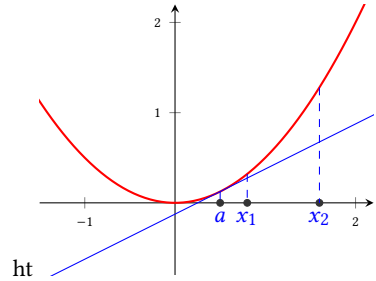

%% file: appendix_rq1_tab.tex
\begin{table*}[thb]
\centering
\caption{Similarity measures (RBO) between ranked lists according to various prioritization metrics. The first column specifies the model, dataset and accuracy on the test dataset. The second column specifies what transformations are used for the experiment. The third column norm measures the average $\ell_2$-norm of all transformations. The rest of the columns measure the similarity between pairs of metric scores. \colorbox{blue!10}{We omit the confidence interval when it is $<0.005$}.}\label{tab:rq1-c}
\begin{adjustbox}{width=1\linewidth,center}
\begin{tabular}{l|l|c|c|c|c|c|c|c|c}
\toprule
\bf{Data} & \bf{Transfor-} & \multirow{3}{*}{\bf Norm} & \multirow{3}{*}{\bf FM\&FL} & \multirow{3}{*}{\bf FM\&BM} & \multirow{3}{*}{\bf FL\&BL} & \multirow{3}{*}{\bf FL\&BD} & \multirow{3}{*}{\bf FL\&BD-F} & \multirow{3}{*}{\bf FL\&ATS} & \multirow{3}{*}{\bf FL\&NC} \\ 
\bf Model, & \bf mation & & & & & & & & \\
\bf Accuracy & & & & & & & & & \\ \hline \hline
MNIST & SD & 0.65  & 0.89  & 0.95  &  0.94 & 0.57 & 0.57  &  0.74 &  0.57 \\
\cline{2-10}
LeNet-1 & Natural & $2.68\pm 0.02$ & 0.94  & 0.84  & 0.75  & 0.76  & 0.77  & 0.81  & 0.67   \\ \cline{2-10}
97.32\% & Mixed  & $1.77\pm 0.01$  &  0.93 & 0.87 & 0.82   & 0.71  & 0.72  &  0.77 & 0.66  \\ \cline{1-10}

MNIST & SD & 0.65 & 0.89  & 0.96  & 0.94  & 0.57 & 0.59 & 0.67  &  0.56 \\
\cline{2-10}
LeNet-5 & Natural & $2.68\pm 0.02$ & 0.95  &  0.84 & 0.76  & 0.77 & 0.79  & 0.81  &  0.66  \\ \cline{2-10}
99.01\% & Mixed  & $1.77\pm 0.01$  & 0.93 & 0.88  & 0.83  & 0.72  & 0.74  & 0.76  &  0.64 \\ \cline{1-10}

SVHN & SD & 0.11  & 0.90 & 0.95 & 0.93  & 0.54 & 0.55 & 0.58   & 0.54 \\
\cline{2-10}
ResNet-9 & Natural & $8.60 \pm 0.02$  & 0.95  & 0.57 &  0.55 & 0.71  & 0.78  & 0.76  &  0.72  \\ \cline{2-10}
95.90\% & Mixed  &  $4.22\pm 0.01$  & 0.94  & 0.68  & 0.66  & 0.69  & 0.72  & 0.72  & 0.70  \\ \cline{1-10}

SVHN & SD &  0.11 & 0.88  & 0.95 & 0.94  & 0.56 & 0.57 & 0.56   &  0.56\\
\cline{2-10}
VGG & Natural &  $8.60\pm 0.02$ & 0.94  & 0.61  & 0.60  &  0.71 &  0.76 & 0.70  &  0.70  \\ \cline{2-10}
95.19\% & Mixed  & $4.22\pm 0.01$   & 0.92  & 0.72  & \hfil 0.70  &   0.67 & 0.70 & 0.67  &   0.67\\ \cline{1-10}

CIFAR10, & SD & 0.11 & 0.90& 0.90 & 0.90 &0.56 &0.57 & 0.58 & 0.56\\\cline{2-10}
ResNet-18, & Natural & $11.42\pm 0.04$ & 0.95 & 0.57 & 0.56 & 0.76 & 0.80 & 0.76  & 0.71 \\
\cline{2-10}
92.29\% & Mixed  & $5.67\pm 0.02$ & 0.95 & 0.68 & 0.66 & 0.73& 0.75& 0.73 & 0.70\\ \cline{1-10}

CIFAR10, & SD & 0.11 & 0.91& 0.93 & 0.92 & 0.56 & 0.58 & 0.56 & 0.56 \\\cline{2-10}
VGG, & Natural & $11.42\pm 0.04$ & 0.96& 0.57 & 0.55 & 0.75 & 0.79 & 0.71  & 0.73 \\
\cline{2-10}
91.48\% & Mixed  & $5.67\pm 0.02$ & 0.95 & 0.68 & 0.66 & 0.71 & 0.73 & 0.69 & 0.70\\ \cline{1-10}

CIFAR100 & SD &  0.11 &  0.85 &  0.88 & 0.88 & 0.55 & 0.57 & -- & 0.55   \\\cline{2-10}
ResNet-18 &  Natural & $11.69\pm 0.04$ & 0.89 & 0.59  & 0.56 & 0.76 & 0.79 & --  & 0.75 \\ \cline{2-10}
70.45\% & Mixed &  $5.80\pm 0.03$ & 0.89 & 0.69 & 0.66 & 0.73 & 0.75  & -- & 0.73 \\\cline{1-10}

CIFAR100 & SD & 0.11 & 0.91 & 0.90  & 0.90 & 0.56 & 0.58 & -- &  0.56  \\\cline{2-10}
VGG &  Natural & $11.69 \pm 0.04$ & 0.95 & 0.59 & 0.57  & 0.75 & 0.78 & --  & 0.72\\ \cline{2-10}
67.73\% & Mixed & $5.80\pm 0.03$  & 0.95 & 0.70  & 0.67 &  0.72 & 0.73 & -- & 0.70 \\
\bottomrule
\end{tabular}
\end{adjustbox}
\end{table*}

%% file: appendix_rq2_tab.tex
\begin{table*}
\begin{center}
\caption{Adaptive standard testing of each metric: the first column specifies the data, model, clean accuracy, and underlying data transformations. The second column specifies which metric is used for testing. The rest of the columns are the accuracy of tests prioritized by each metric during each iteration in adaptive testing. We include the confidence interval when the result from diversity metrics is close to our best directed-metric result.}\label{tab:rq2-c1}
\begin{adjustbox}{width=\linewidth,center}
\begin{tabular}{C{2.8cm}|L{1.2cm}|C{1.4cm}C{1.4cm}C{1.4cm}C{1.4cm}C{1.4cm}}
\toprule
\multirow{3}{*}{\shortstack[c]{\bf Data, Model, \\ \bf Accuracy, \\ \bf Transformations}} & \multirow{3}{*}{\bf Fitness} & \multirow{3}{*}{\bf Iter 1} & \multirow{3}{*}{\bf Iter 2} & \multirow{3}{*}{\bf Iter 3} & \multirow{3}{*}{\bf Iter 4} & \multirow{3}{*}{\bf Iter 5} \\ 
& & & & & & \\
& & & & & & \\ \hline \hline
&  FM &  6.66\%&  0.49\% &  0.36\% &  0.36\% &  0.36\%\\
\cline{2-7}
 &  FL  & 10.20\%&  0.11\%&  0.07\%&  0.07\% &  0.07\% \\ \cline{2-7}
MNIST  &  BM &  28.46\% &  10.27\% &  4.51\% &  2.58\% &  1.87\% \\ \cline{2-7}
LeNet-1 &  BL &  41.42\% &  16.36\% &  7.58\% &  4.23\% &  2.54\% \\ \cline{2-7}
97.32\% &  BD &  25.5\%&  16.2\% &  14.27\% &  13.01\% &  12.30\%\\ \cline{2-7}
Natural &  BD-F &  18.05\%&  7.04\% &  4.69\% &  3.88\% &  3.49\%\\ \cline{2-7}
 &  ATS &  29.56\% &  19.86\% &  17.31\% &  16.42\%&  16.24\%\\ \cline{2-7}
 &  NC &  78.83\%&  76.90\% &  74.52\% &  72.64\% &  71.71\% \\ \cline{1-7}
&  FM &  4.93\%&  0.08\% &  0.01\% &  0.0\% &  0.0\%\\ \cline{2-7}
&  FL  & 4.97\%&  0.1\%&  0.01\%&  0.0\% &  0.0\% \\ \cline{2-7}
MNIST &  BM &  28.63\% &  10.13\% &  4.33\% &  2.04\% &  1.37\% \\ \cline{2-7}
LeNet-5&  BL &  37.63\% &  13.27\% &  5.19\% &  2.43\% &  1.43\% \\
\cline{2-7}
99.01\% &  BD &  25.43\%&  11.51\% &  9.99\% &  9.18\% &  9.13\%\\ \cline{2-7}
Natural &  BD-F &  14.22\%&  4.06\% &  2.52\% &  2.11\% &  1.88\%\\ \cline{2-7}
 &  ATS &  27.74\%&  19.97\% &  17.6\% &  16.61\% &  16.33\%\\ \cline{2-7}
 &  NC &  80.51\%&  79.72\% &  78.83\% &  78.09\% &  77.66\% \\ \cline{1-7}
&  FM &  20.71\%&  1.95\% & 0.38\% &  0.29\% &  0.26\%\\
\cline{2-7}
&  FL  &  20.91\%&  2.19\%&  0.47\%&  0.33\% &  0.29\% \\
\cline{2-7}
SVHN &  BM &  80.72\% &  57.59\% &  41.37\% &  31.87\% &  24.21\% \\
\cline{2-7}
ResNet-9&  BL &  81.29\% &  59.57\% &  43.64\% &  33.72\% &  25.60\% \\
\cline{2-7}
95.90\%   &  BD &  36.16\%&  13.98\% &  10.23\% &  9.82\% &  9.73\%\\
\cline{2-7}
Natural &  BD-F &  24.93\%&  4.58\% &  1.77\% &  1.29\% &  1.09\%\\
\cline{2-7}
 &  ATS &  30.39\%&  19.52\% &  17.06\% &  15.84\% &  15.47\%\\
\cline{2-7}
 &  NC &  39.75\%&  15.17\% &  10.51\% &  9.94\% &  9.92\% \\
\cline{1-7}
 &  FM &  18.12\%&  4.28\% &  2.79\% &  2.34\% &  2.05\%\\
\cline{2-7}
 &  FL  &  18.15\%&  4.79\%&  3.33\%&  2.83\% &  2.50\% \\
\cline{2-7}
SVHN &  BM &  74.77\% &  54.19\% &  38.43\% &  30.36\% &  24.23\% \\
\cline{2-7}
VGG &  BL &  75.52\% &  52.11\% &  37.04\% &  28.30\% &  21.66\% \\
\cline{2-7}
95.19\% &  BD &  28.96\%&  18.57\% &  19.45\% &  19.52\% &  19.51\%\\
\cline{2-7}
Natural &  BD-F &  20.64\%&  6.53\% &  4.71\% &  4.11\% &  3.71\%\\
\cline{2-7}
 &  ATS &  33.05\%&  22.69\% &  19.88\% &  18.70\% &  17.91\%\\
\cline{2-7}
 &  NC &  33.80\%&  23.40\% &  21.59\% &  20.95\% &  20.70\% \\
\cline{1-7}
\end{tabular}
\end{adjustbox}
\end{center}
\end{table*}

\begin{table*}
\begin{center}
 
\caption*{Table (Continued)}
\begin{adjustbox}{width=\linewidth,center}
\begin{tabular}{C{2.8cm}|L{1.2cm}|C{1.4cm}C{1.4cm}C{1.4cm}C{1.4cm}C{1.4cm}}
\cline{1-7}
 &  FM &  10.78\% &  1.02\% &  0.42\% &  0.25\% &  0.20\%  \\
\cline{2-7}
 &  FL  &  10.92\% &  1.17\% &  0.47\% &  0.27\% &  0.20\%  \\
\cline{2-7}
  CIFAR10, &  BM &  71.45\% &  55.24\% &  41.98\% &  33.90\% &  27.93\%  \\
\cline{2-7}
  ResNet-18, &  BL &  78.63\% &  63.79\% &  50.71\% &  41.79\% &  35.42\%  \\
\cline{2-7}
 92.29\% &  BD &  24.41\% &  13.38\% &  11.75\% &  11.50\% &  11.45\%  \\
\cline{2-7}
 Natural &  BD-F &  17.00\% &  4.56\% &  2.41\% &  1.74\% &  1.57\% \\
\cline{2-7}
  &  ATS &  29.49\% &  21.46\% &  19.92\% &  18.61\% &  17.95\% \\
\cline{2-7}
 &  NC  &  40.41\% &  30.49\% &  26.20\% &  24.53\% &  23.50\% \\
\cline{1-7}
 &  FM &  8.43\% &  1.26\% &  0.56\% &  0.44\% &  0.32\%  \\
\cline{2-7}
 &  FL  &  8.46\% &  1.34\% &  0.68\% &  0.56\% &  0.47\%  \\
\cline{2-7}
CIFAR10 &  BM &  74.82\% &  59.75\% &  47.60\% &  39.35\% &  32.23\%  \\
\cline{2-7}
VGG &  BL &  74.96\% &  60.4\% &  48.92\% &  40.27\% &  33.51\%  \\
\cline{2-7}
91.48\% &  BD &  16.85\% &  9.71\% &  9.53\% &  9.35\% &  9.34\%  \\
\cline{2-7}
Natural &  BD-F &  11.56\% &  3.23\% &  2.05\% &  1.61\% &  1.28\% \\
\cline{2-7}
  &  ATS &  29.49\% &  21.46\% &  19.92\% &  18.61\% &  17.95\% \\
\cline{2-7}
 &  NC  &  24.32\% &  15.14\% &  12.82\% &  11.82\% &  11.28\% \\
\cline{1-7}
 &  FM &  3.32\% &  0.27\% &  $0.10\pm 0.06$\% &  $0.03\pm 0.03$\% &  $0.02\pm 0.02$\%  \\
\cline{2-7}
CIFAR100 &  FL  &  3.88\% &  0.29\% &  $0.10\pm 0.06$\% &  $0.06\pm 0.05$\% &  $0.05\pm 0.04$\%  \\
\cline{2-7}
ResNet-18 &  BM &  50.19\% &  35.72\% &  26.21\% &  19.65\% &  14.97\%  \\
\cline{2-7}
70.45\%  &  BL &  50.03\% &  35.84\% &  26.46\% &  19.81\% &  15.46\%  \\
\cline{2-7}
 Natural &  BD &  10.54\% &  1.67\% &  0.98\% &  1.01\% &  0.96\%  \\
\cline{2-7}
 &  BD-F &  5.69\% &  0.62\% &  $0.36\pm 0.11$\% &  $0.20\pm 0.09$\% &  $0.18\pm 0.08$\% \\
\cline{2-7}
 &  NC &  12.95\% &  3.88\% &  2.37\% &  1.94\% &  1.80\% \\
 \cline{1-7}
 &  FM &  3.10\% &  0.24\% &   0.05\% &  0.01\% &  0.01\%  \\
\cline{2-7}
 CIFAR100, &  FL  &  3.18\% &  0.23\% &  0.04\% &  0.02\% &  0.01\%  \\
\cline{2-7}
 VGG &  BM &  47.61\% &  33.49\% &  23.82\% &  17.72\% &  13.34\%  \\
\cline{2-7}
 67.73\% &  BL &  47.39\% &  33.54\% &  23.95\% &  18.11\% &  13.89\%  \\
\cline{2-7}
Natural &  BD &  7.33\% &  1.29\% &  0.92\% &  0.82\% &  0.77\% \\
\cline{2-7}
 & BD-F &  4.77\% &  0.63\% &  0.33\% &  0.28\% &  0.23\% \\
\cline{2-7}
 &  NC &  13.66\% &  7.21\% &  5.66\% &  4.95\% &  4.51\% \\
 \cline{1-7}
 &  FM &  3.10\% &  0.24\% &   0.05\% &  0.01\% &  0.01\%  \\
\cline{2-7}
 CIFAR100, &  FL  &  3.18\% &  0.23\% &  0.04\% &  0.02\% &  0.01\%  \\
\cline{2-7}
 VGG &  BM &  47.61\% &  33.49\% &  23.82\% &  17.72\% &  13.34\%  \\
\cline{2-7}
 67.73\% &  BL &  47.39\% &  33.54\% &  23.95\% &  18.11\% &  13.89\%  \\
\cline{2-7}
Natural &  BD &  7.33\% &  1.29\% &  0.92\% &  0.82\% &  0.77\% \\
\cline{2-7}
 & BD-F &  4.77\% &  0.63\% &  0.33\% &  0.28\% &  0.23\% \\
\cline{2-7}
 &  NC &  13.66\% &  7.21\% &  5.66\% &  4.95\% &  4.51\% \\
\bottomrule
\end{tabular}
\end{adjustbox}
\end{center}
\end{table*}

%% file: appendix_rq3_tab1.tex
\begin{table*}
\begin{center}
\caption{Adaptive standard testing of each metric: the first column specifies the data, model, clean accuracy, and underlying data transformations. The second column specifies which metric is used for testing. The rest of the columns are the accuracy of tests prioritized by each metric during each iteration in adaptive testing. We include the confidence interval when the result from diversity metrics is close to our best directed-metric result.}\label{tab:rq2-c2}
\begin{adjustbox}{width=\linewidth,center}
\begin{tabular}{C{2.8cm}|L{1.2cm}|C{1.4cm}C{1.4cm}C{1.4cm}C{1.4cm}C{1.4cm}}
\toprule
\multirow{3}{*}{\shortstack[c]{\bf Data Model \\ \bf Accuracy \\ \bf Transformations}} & \multirow{3}{*}{\bf Fitness} & \multirow{3}{*}{\bf Iter 1} & \multirow{3}{*}{\bf Iter 2} & \multirow{3}{*}{\bf Iter 3} & \multirow{3}{*}{\bf Iter 4} & \multirow{3}{*}{\bf Iter 5} \\ 
& & & & & & \\
& & & & & & \\ \hline \hline
 &  FM &  96.35\%&  94.93\% &  92.89\% &  90.20\% &  86.62\%\\
\cline{2-7}
 &  FL  &  96.38\%&  95.06\%&  93.02\%&  90.37\% &  86.98\% \\
\cline{2-7}
  MNIST &  BM &  70.36\% &  15.45\% &  2.47\% &  0.28\% &  0.03\% \\
\cline{2-7}
  LeNet-1 &  BL &  33.02\% &  0.65\% &  0.0\% &  0.0\% &  0.0\% \\
\cline{2-7}
 97.32\% &  BD &  97.32\%&  97.27\% &  97.14\% &  97.02\% &  96.90\%\\
\cline{2-7}
 SD &  BD-F &  97.21\%&  97.19\% &  97.08\% &  96.93\% &  96.67\%\\
\cline{2-7}
 &  ATS &  97.19\%&  96.61\% &  95.67\% &  94.62\% &  93.0\%\\
\cline{2-7}
 &  NC &  97.27\%&  97.24\% &  97.26\% &  97.19\% &  97.26\% \\
\cline{1-7}
 &  FM &  98.32\%&  97.42\% &  96.14\% &  94.36\% &  91.74\%\\
\cline{2-7}
 &  FL  &  98.34\%&  97.41\%&  96.17\%&  94.43\% &  91.89\% \\
\cline{2-7}
  MNIST &  BM &  69.15\% &  13.62\% &  1.51\% &  0.05\% &  0.0\% \\
\cline{2-7}
  LeNet-5 &  BL &  42.06\% &  1.39\% &  0.01\% &  0.0\% &  0.0\% \\
\cline{2-7}
 99.01\% &  BD &  98.89\%&  98.84\% &  98.75\% &  98.65\% &  98.62\%\\
\cline{2-7}
 SD &  BD-F &  98.91\%&  98.77\% &  98.55\% &  98.21\% &  97.66\%\\
\cline{2-7}
 &  ATS &  98.90\%&  98.50\% &  97.90\% &  96.83\% &  95.38\%\\
\cline{2-7}
 &  NC &  98.89\%&  98.89\% &  98.85\% &  98.82\% &  98.84\% \\
\cline{1-7}
 &  FM &  95.61\%&  95.31\% &  95.00\% &  94.69\% &  94.30\%\\
\cline{2-7}
 &  FL  &  95.62\%&  95.32\%&  95.03\%&  94.74\% &  94.34\% \\
\cline{2-7}
  SVHN &  BM &  89.23\% &  77.71\% &  63.88\% &  50.65\% &  39.01\% \\
\cline{2-7}
  ResNet-9 &  BL &  86.45\% &  71.06\% &  54.72\% &  40.96\% &  29.85\% \\
\cline{2-7}
 95.90\% &  BD &  95.88\%&  95.87\% &  95.87\% &  95.89\% &  95.87\%\\
\cline{2-7}
 SD &  BD-F &  95.87\%&  95.86\% &  95.83\% &  95.82\% &  95.77\%\\
\cline{2-7}
 &  ATS &  95.89\%&  95.81\% &  95.77\% &  95.70\% &  95.63\%\\
\cline{2-7}
 &  NC &  95.87\%&  95.91\% &  95.90\% &  95.89\% &  95.89\% \\
\cline{1-7}
 &  FM &  94.93\%&  94.57\% &  94.21\% &  93.86\% &  93.56\%\\
\cline{2-7}
 &  FL  &  94.94\%&  94.58\%&  94.23\%&  93.87\% &  93.57\% \\
\cline{2-7}
  SVHN &  BM &  85.38\% &  73.49\% &  62.97\% &  53.38\% &  44.24\% \\
\cline{2-7}
  VGG &  BL &  84.88\% &  71.77\% &  59.85\% &  49.03\% &  39.40\% \\
\cline{2-7}
 95.19\% &  BD &  95.18\%&  95.16\% &  95.15\% &  95.08\% &  95.06\%\\
\cline{2-7}
 SD &  BD-F &  95.18\%&  95.17\% &  95.15\% &  95.09\% &  95.01\%\\
\cline{2-7}
 &  ATS &  95.21\%&  95.17\% &  94.95\% &  94.91\% &  94.87\%\\
\cline{2-7}
 &  NC &  95.17\%&  95.16\% &  95.12\% &  95.08\% &  95.03\% \\
\bottomrule
\end{tabular}
\newline\newline 
\end{adjustbox}
\end{center}
\end{table*}

\begin{table*}
\begin{center}
\caption*{Table (Continued)}
\begin{adjustbox}{width=.95\linewidth,center}
\begin{tabular}{C{2.8cm}|L{1.2cm}|C{1.4cm}C{1.4cm}C{1.4cm}C{1.4cm}C{1.4cm}}
\cline{1-7}
 &  FM &  91.40\% &  90.55\% &  89.31\% &  88.24\% &  87.11\%  \\
\cline{2-7}
 &  FL  &  91.44\% &  90.59\% &  89.33\% &  88.33\% &  87.26\%  \\
\cline{2-7}
   CIFAR10 &  BM &  67.76\% &  42.21\% &  26.08\% &  15.82\% &  9.53\%  \\
\cline{2-7}
   ResNet-18 &  BL &  64.75\% &  37.16\% &  20.80\% &  10.94\% &  6.08\%  \\
\cline{2-7}
  92.29\% &  BD &  92.31\% &  92.35\% &  92.20\% &  92.18\% &  92.12\%  \\
\cline{2-7}
  SD &  BD-F &  92.27\% &  92.22\% &  92.20\% &  91.98\% &  91.90\% \\
\cline{2-7}
  &  ATS &  92.27\% &  92.32\% &  91.70\% &  91.51\% &  91.23\% \\
\cline{2-7}
 &  NC  &  92.25\% &  92.21\% &  92.26\% &  92.18\% &  92.20\% \\
\cline{1-7}
 &  FM &  90.90\% &  90.05\% &  89.40\% &  88.61\% &  87.76\%  \\
\cline{2-7}
 &  FL  &  90.90\% &  90.05\% &  89.43\% &  88.63\% &  87.74\%  \\
\cline{2-7}
  CIFAR10 &  BM &  70.45\% &  50.69\% &  36.69\% &  26.54\% &  18.94\%  \\
\cline{2-7}
  VGG &  BL &  69.54\% &  48.64\% &  34.26\% &  24.22\% &  16.55\%  \\
\cline{2-7}
   91.48\% &  BD &  91.46\% &  91.61\% &  91.50\% &  91.30\% &  91.25\%  \\
\cline{2-7}
  SD &  BD-F &  91.47\% &  91.51\% &  91.40\% &  91.18\% &  90.93\% \\
\cline{2-7}
  &  ATS &  91.46\% &  91.37\% &  91.13\% &  91.31\% &  91.00\% \\
\cline{2-7}
 &  NC  &  91.52\% &  91.52\% &  91.55\% &  91.53\% &  91.35\% \\
\cline{1-7}
 &  FM &  68.18\% &  65.77\% &  63.47\% &  61.30\% &  59.09\%  \\
\cline{2-7}
 CIFAR100 &  FL  &  68.58\% &  66.53\% &  64.52\% &  62.60\% &  60.62\%  \\
\cline{2-7}
ResNet-18 &  BM &  36.73\% &  13.77\% &  4.40\% &  1.51\% &  0.57\%  \\
\cline{2-7}
  70.45\%  &  BL &  33.07\% &  11.27\% &  3.53\% &  1.22\% &  0.42\%  \\
\cline{2-7}
 SD &  BD &  70.42\% &  70.33\% &  70.39\% &  70.42\% &  70.25\%  \\
\cline{2-7}
 &  BD-F &  70.30\% &  70.08\% &  69.80\% &  69.60\% &  69.04\% \\
\cline{2-7}
 &  NC &  70.51\% &  70.51\% &  70.29\% &  70.17\% &  70.14\% \\
 \cline{1-7}
 &  FM &  66.05\% &  64.42\% &  62.90\% &  61.48\% &  59.97\%  \\
\cline{2-7}
 CIFAR100 &  FL  &  66.09\% &  64.54\% &  63.01\% &  61.65\% &  60.18\%  \\
\cline{2-7}
  VGG &  BM &  40.32\% &  21.43\% &  11.39\% &  5.69\% &  3.02\%  \\
\cline{2-7}
  67.73\%  &  BL &  36.42\% &  17.75\% &  8.39\% &  3.86\% &  2.06\%  \\
\cline{2-7}
 SD &  BD &  67.73\% &  67.84\% &  67.78\% &  67.68\% &  67.60\%  \\
\cline{2-7}
 &  BD-F &  67.64\% &  67.45\% &  67.30\% &  67.15\% &  67.16\% \\
\cline{2-7}
 &  NC &  67.75\% &  67.78\% &  67.67\% &  67.73\% &  67.67\% \\
\bottomrule
\end{tabular}
\newline\newline 
\end{adjustbox}
\end{center}
\end{table*}

%% file: appendix_rq3_tab2.tex
\begin{table*}
\begin{center}
\caption{Adaptive standard testing of each metric: the first column specifies the data, model, clean accuracy, and underlying data transformations. The second column specifies which metric is used for testing. The rest of the columns are the accuracy of tests prioritized by each metric during each iteration in adaptive testing. We include the confidence interval when the result from diversity metrics is close to our best directed-metric result. Notice that we do not include the result of NC for SVHN because it is too slow to finish the iterations (over $30$ hours).}\label{tab:rq2-c3}
\begin{adjustbox}{width=\linewidth,center}
\begin{tabular}{C{2.8cm}|L{1.2cm}|C{1.4cm}C{1.4cm}C{1.4cm}C{1.4cm}C{1.4cm}}
\toprule
\multirow{3}{*}{\shortstack[c]{\bf Data Model \\ \bf Accuracy \\ \bf Transformations}} & \multirow{3}{*}{\bf Fitness} & \multirow{3}{*}{\bf Iter 1} & \multirow{3}{*}{\bf Iter 2} & \multirow{3}{*}{\bf Iter 3} & \multirow{3}{*}{\bf Iter 4} & \multirow{3}{*}{\bf Iter 5} \\ 
& & & & & & \\
& & & & & & \\ \hline \hline
    &     FM &     10.06\%&     0.37\% &     0.0\% &     0.0\% &     0.0\%\\
 \cline{2-7}
    &     FL  &     10.44\%&     0.39\%&     0.02\%&     0.0\% &     0.0\% \\
 \cline{2-7}
     &     BM &     35.94\% &     12.81\% &     5.79\% &     3.28\% &     2.16\% \\
 \cline{2-7}
  MNIST   &     BL &     49.18\% &     20.23\% &     10.11\% &     5.94\% &     3.30\% \\
 \cline{2-7}
  LeNet-1   &     MM &     10.05\% &     0.26\% &     0.01\% &     0.0\% &     0.0\% \\
 \cline{2-7}
  97.32\%   &     ML &     10.38\% &     0.16\% &     0.0\% &     0.0\% &     0.0\% \\
 \cline{2-7}
    Mixed &     BD &     27.68\%&     16.50\% &     14.72\% &     14.08\% &     12.57\%\\
 \cline{2-7}
     &     BD-F &     21.09\%&     7.60\% &     4.74\% &     3.67\% &     3.43\%\\
 \cline{2-7}
    &     ATS &     34.41\%&     22.92\% &     19.78\% &     17.56\% &     16.70\%\\
 \cline{2-7}
    &     NC &     78.52\%&     78.38\% &     78.10\% &     78.02\% &     78.07\% \\
  \cline{1-7}
    &     FM &     8.47\%&     0.07\% &     0.0\% &     0.0\% &     0.0\%\\
 \cline{2-7}
    &     FL  &     8.56\%&     0.09\%&     0.0\%&     0.0\% &     0.0\% \\
 \cline{2-7}
     &     BM &     40.23\% &     12.43\% &     5.12\% &     2.17\% &     1.27\% \\
 \cline{2-7}
  MNIST   &     BL &     48.78\% &     18.41\% &     6.98\% &     3.67\% &     1.77\% \\
 \cline{2-7}
  LeNet-5   &     MM &     8.47\% &     0.07\% &     0.0\% &     0.0\% &     0.0\% \\
 \cline{2-7}
  99.01\%   &     ML &     8.52\% &     0.05\% &     0.0\% &     0.0\% &     0.0\% \\
 \cline{2-7}
    Mixed &     BD &     36.45\%&     9.26\% &     8.99\% &     8.92\% &     8.85\%\\
 \cline{2-7}
     &     BD-F &     18.60\%&     4.88\% &     2.94\% &     2.57\% &     2.26\%\\
 \cline{2-7}
    &     ATS &     31.08\%&     21.58\% &     19.01\% &     18.17\% &     16.82\%\\
 \cline{2-7}
    &     NC &     88.81\%&     87.93\% &     87.39\% &     87.13\% &     86.95\% \\
  \cline{1-7}
    &     FM &     15.04\%&     2.65\% &     0.56\% &     0.04\% &     0.0\%\\
 \cline{2-7}
    &     FL  &     15.31\%&     2.89\%&     0.68\%&     0.08\% &     0.02\% \\
 \cline{2-7}
     &     BM &     79.26\% &     57.53\% &     38.51\% &     26.85\% &     18.97\% \\
 \cline{2-7}
  SVHN   &     BL &     76.82\%&     51.28\% &     31.30\% &     18.42\% &     11.41\% \\
 \cline{2-7}
  ResNet-9   &     MM &     14.58\% &     1.62\% &     0.12\% &     0.02\% &     0.0\% \\
 \cline{2-7}
  95.90\%   &     ML &     14.54\% &     1.66\% &     0.13\% &     0.0\% &     0.0\% \\
 \cline{2-7}
    Mixed &     BD &     34.93\%&     17.16\% &     11.24\% &     10.20\% &     10.17\%\\
 \cline{2-7}
     &     BD-F &     18.53\%&     5.16\% &     2.05\% &     1.10\% &     0.86\%\\
 \cline{2-7}
    &     ATS &     30.88\%&     19.96\% &     17.83\% &     16.81\% &     16.29\%\\
\bottomrule
\end{tabular}
\newline\newline 
\end{adjustbox}
\end{center}
\end{table*}

\begin{table*}
\begin{center}
\caption*{Table (Continued)}
\begin{adjustbox}{width=\linewidth,center}
\begin{tabular}{C{2.8cm}|L{1.2cm}|C{1.4cm}C{1.4cm}C{1.4cm}C{1.4cm}C{1.4cm}}
  \cline{1-7}
    &     FM &     16.29\%&     4.93\% &     2.70\% &     2.14\% &     1.85\%\\
 \cline{2-7}
    &     FL  &     16.35\%&     5.42\%&     3.24\%&     2.57\% &     2.23\% \\
 \cline{2-7}
     &     BM &     74.12\% &     52.57\% &     35.21\% &     23.53\% &     15.73\% \\
 \cline{2-7}
  SVHN   &     BL &     75.02\% &     53.71\% &     35.45\% &     22.37\% &     13.86\% \\
 \cline{2-7}
  VGG   &     MM &     16.76\% &     1.33\% &     0.10\% &     0.00\% &     0.00\% \\
 \cline{2-7}
  95.19\%   &     ML &     16.63\% &     1.36\% &     0.10\% &     0.01\% &     0.00\% \\
 \cline{2-7}
    Mixed &     BD &     27.64\%&     19.58\% &     19.22\% &     19.49\% &     19.56\%\\
 \cline{2-7}
     &     BD-F &     18.75\%&     7.21\% &     4.86\% &     4.08\% &     3.58\%\\
 \cline{2-7}
    &     ATS &     33.97\%&     22.78\% &     19.80\% &     18.49\% &     17.46\%\\
  \cline{1-7}
    &     FM &     8.01\%&     0.99\% &     0.29\% &     0.16\% &     0.07\%\\
 \cline{2-7}
    &     FL  &     8.14\%&     1.06\%&     0.33\%&     0.17\% &     0.08\% \\
 \cline{2-7}
     &     BM &     58.85\% &     33.99\% &     21.55\% &     14.90\% &     10.88\% \\
 \cline{2-7}
  CIFAR10   &     BL &     54.01\% &     25.60\% &     12.16\% &     6.22\% &     3.39\% \\
 \cline{2-7}
  ResNet-18   &     MM &     6.79\% &     0.30\% &     0.05\% &     0.01\% &     0.00\% \\
 \cline{2-7}
  92.29\%   &     ML &     6.69\% &     0.30\% &     0.05\% &     0.00\% &     0.00\% \\
 \cline{2-7}
    Mixed &     BD &     22.02\%&     12.73\% &     12.15\% &     12.20\% &     12.36\%\\
 \cline{2-7}
     &     BD-F &     13.91\%&     4.13\% &     2.37\% &     1.77\% &     1.41\%\\
 \cline{2-7}
    &     ATS &     31.02\% &     20.91\% &     18.58\% &     18.06\% &     17.81\% \\
 \cline{2-7}
    &     NC &     35.87\%&     31.98\% &     31.20\% &     31.04\% &     30.86\% \\
  \cline{1-7}
    &     FM &     6.17\%&     1.40\% &     0.91\% &     0.70\% &     0.55\%\\
 \cline{2-7}
    &     FL  &     6.22\%&     1.52\%&     1.05\%&     0.88\% &     0.79\% \\
 \cline{2-7}
     &     BM &     64.06\% &     43.81\% &     30.27\% &     21.07\% &     14.81\% \\
 \cline{2-7}
  CIFAR10   &     BL &     63.13\% &     40.55\% &     25.21\% &     15.81\% &     9.46\% \\
 \cline{2-7}
  VGG   &     MM &     4.86\% &     0.48\% &     0.11\% &     0.01\% &     0.00\% \\
 \cline{2-7}
  91.48\%   &     ML &     4.84\% &     0.50\% &     0.11\% &     0.02\% &     0.00\% \\
 \cline{2-7}
    Mixed &     BD &     15.90\%&     9.75\% &     9.59\% &     9.61\% &     9.61\%\\
 \cline{2-7}
     &     BD-F &     9.38\%&     3.38\% &     2.60\% &     2.27\% &     2.02\%\\
 \cline{2-7}
    &     ATS &     30.83\% &     22.69\% &     20.51\% &     19.59\% &     20.01\% \\
 \cline{2-7}
    &     NC &     20.74\%&     14.07\% &     13.07\% &     12.73\% &     12.58\% \\
      \cline{1-7}
    &     FM &     2.17\%&     0.22\% &     0.03\% &     0.00\% &     0.00\%\\
 \cline{2-7}
    &     FL  &     2.46\%&     0.28\%&     0.08\%&     0.04\% &     0.01\% \\
 \cline{2-7}
     &     BM &     28.91\% &     15.24\% &     10.62\% &     7.69\% &     5.19\% \\
 \cline{2-7}
  CIFAR100   &     BL &     23.54\% &     6.16\% &     2.26\% &     1.13\% &     0.87\% \\
 \cline{2-7}
  ResNet-18   &     MM &     1.56\% &     0.01\% &     0.00\% &     0.00\% &     0.00\% \\
 \cline{2-7}
  70.45\%   &     ML &     1.67\% &     0.01\% &     0.00\% &     0.00\% &     0.00\% \\
 \cline{2-7}
    Mixed &     BD &     7.84\%&     1.76\% &     1.15\% &     1.12\% &     1.07\%\\
 \cline{2-7}
     &     BD-F &     4.50\%&     0.72\% &     0.40\% &     0.30\% &     0.27\%\\
 \cline{2-7}
 \cline{2-7}
    &     NC &     10.06\%&     3.90\% &     3.01\% &     2.68\% &     2.65\% \\
\bottomrule
\end{tabular}
\newline\newline 
\end{adjustbox}
\end{center}
\end{table*}

\begin{table*}
\begin{center}
\caption*{Table (Continued)}
\begin{adjustbox}{width=\linewidth,center}
\begin{tabular}{C{2.8cm}|L{1.2cm}|C{1.4cm}C{1.4cm}C{1.4cm}C{1.4cm}C{1.4cm}}
  \cline{1-7}
    &     FM &     1.94\%&     0.19\% &     0.05\% &     0.01\% &     0.01\%\\
 \cline{2-7}
    &     FL  &     1.97\%&     0.22\%&     0.05\%&     0.02\% &     0.01\% \\
 \cline{2-7}
     &     BM &     36.23\% &     19.18\% &     12.05\% &     7.70\% &     5.46\% \\
 \cline{2-7}
  CIFAR100   &     BL &     30.14\% &     12.30\% &     5.62\% &     3.48\% &     1.99\% \\
 \cline{2-7}
  VGG   &     MM &     1.63\% &     0.07\% &     0.02\% &     0.01\% &     0.00\% \\
 \cline{2-7}
  67.73\%   &     ML &     1.66\% &     0.09\% &     0.02\% &     0.00\% &     0.00\% \\
 \cline{2-7}
    Mixed &     BD &     5.35\%&     1.11\% &     0.85\% &     0.79\% &     0.80\%\\
 \cline{2-7}
     &     BD-F &     4.00\%&     0.70\% &     0.43\% &     0.38\% &     0.35\%\\
 \cline{2-7}
 \cline{2-7}
    &     NC &     11.94\%&     6.90\% &     5.83\% &     5.59\% &     5.32\% \\
\bottomrule
\end{tabular}
\newline\newline 
\end{adjustbox}
\end{center}
\end{table*}

%% file: mt-tab.tex
\begin{table*}
\begin{center}
\caption{Adaptive metamorphic testing of forward fitness scores and BD scores from~\citet{diversity-boosting}: the first column specifies the data, model, pseudo-accuracy, and underlying data transformations. The second column specifies which metric is used for testing. The rest of the columns are the pseudo-accuracy of tests prioritized by each metric during each iteration in adaptive testing and the accuracy is relative to pseudo-labels rather than ground-truth labels.}\label{tab:mt-nat}
\begin{adjustbox}{width=\linewidth, center}
\begin{tabular}{C{2.8cm}|L{1.2cm}|C{1.4cm}C{1.4cm}C{1.4cm}C{1.4cm}C{1.4cm}}

\toprule
\multirow{3}{*}{\shortstack[c]{\bf Data Model \\ \bf Pseudo-accuracy \\ \bf Transformations}} & \multirow{3}{*}{\bf Fitness} & \multirow{3}{*}{\bf Iter 1} & \multirow{3}{*}{\bf Iter 2} & \multirow{3}{*}{\bf Iter 3} & \multirow{3}{*}{\bf Iter 4} & \multirow{3}{*}{\bf Iter 5} \\ 
& & & & & & \\
& & & & & & \\ \hline \hline
  &   FM &   10.19\% &   0.22\% &   0.10\% &   0.10\% &   0.09\%  \\
  \cline{2-7}
  MNIST  &   FL  &   10.25\% &   0.23\% &   0.07\% &   0.07\% &   0.07\%   \\
  \cline{2-7}
  LeNet-1  &   BM &   32.56\% &   12.46\% &   4.28\% &   2.48\% &   1.82\%  \\
  \cline{2-7}
  100\%  &   BL  &   45.47\% &   18.79\% &   7.97\% &   4.44\% &   2.81\%   \\
  \cline{2-7}
  Natural &  BD &   25.93\% &   15.78\% &   13.32\% &   12.85\% &   12.62\%   \\
  \cline{2-7}
    &   BD-F &   22.75\% &   7.36\% &   3.51\% &   2.83\% &   2.64\%  \\
  \cline{1-7}
  &   FM  &   8.65\% &   0.10\% &   0.01\% &   0.00\% &   0.00\% \\
  \cline{2-7}
  MNIST  &   FL  &   8.72\% &   0.12\% &   0.01\% &   0.00\% &   0.00\%   \\
  \cline{2-7}
  LeNet-5  &   BM &   32.41\% &   10.84\% &   4.43\% &   2.61\% &   1.41\%  \\
  \cline{2-7}
  100\%  &   BL  &   41.71\% &   14.75\% &   5.34\% &   2.89\% &   1.68\%   \\
  \cline{2-7}
  Natural &  BD &   29.37\% &   12.71\% &   9.81\% &   9.06\% &   8.43\%   \\
  \cline{2-7}
    &   BD-F &   18.53\% &   5.01\% &   2.39\% &   1.90\% &   1.57\%  \\
  \cline{1-7}
   &   FM &   18.47\% &   2.24\% &   0.72\% &   0.51\% &   0.45\%  \\
  \cline{2-7}
  SVHN  &   FL  &   18.68\% &   2.45\% &   0.88\% &   0.67\% &   0.53\%  \\
  \cline{2-7}
  ResNet-9  &  BM &   76.85\% &   54.48\% &   39.45\% &   29.22\% &   21.46\%   \\
  \cline{2-7}
  100\%  &   BL &   78.13\% &   57.59\% &   42.78\% &   32.56\% &   24.09\%  \\
  \cline{2-7}
    Natural &  BD &   37.93\% &   14.76\% &   10.66\% &   10.11\% &   10.02\%   \\
  \cline{2-7}
    &   BD-F &   21.74\% &   4.31\% &   2.07\% &   1.46\% &   1.19\%  \\
  \cline{1-7}
   &   FM &   19.61\% &   4.58\% &   3.19\% &   2.80\% &   2.46\%  \\
  \cline{2-7}
  SVHN  &   FL  &   19.67\% &   4.99\% &   3.70\% &   3.27\% &   2.89\%  \\
  \cline{2-7}
  VGG  &  BM &   72.54\% &   48.12\% &   33.80\% &   24.93\% &   19.14\%   \\
  \cline{2-7}
  100\%  &   BL &   73.66\% &   50.23\% &   35.84\% &   26.55\% &   20.33\%  \\
  \cline{2-7}
    Natural &  BD &   29.92\% &   18.82\% &   19.59\% &   19.76\% &   19.84\%   \\
  \cline{2-7}
    &   BD-F &   21.35\% &   6.48\% &   4.83\% &   4.20\% &   3.80\%  \\
\bottomrule
\end{tabular}
\newline\newline 
\end{adjustbox}
\end{center}
\end{table*}

\begin{table*}
\begin{center}
\caption*{Table (Continued)}
\begin{adjustbox}{width=\linewidth, center}
\begin{tabular}{C{2.8cm}|L{1.2cm}|C{1.4cm}C{1.4cm}C{1.4cm}C{1.4cm}C{1.4cm}}
  \cline{1-7}
    &   FM &   11.43\% &   1.69\% &   0.86\% &   0.60\% &   0.40\%  \\
  \cline{2-7}
  CIFAR10  &   FL  &   11.56\% &   1.83\% &   0.94\% &   0.69\% &   0.53\%  \\
  \cline{2-7}
  ResNet-18  &   BM &   76.11\% &   59.72\% &   46.11\% &   37.77\% &   31.57\%  \\
  \cline{2-7}
  100\%  &   BL  &   76.13\% &   59.95\% &   46.77\% &   38.24\% &   31.83\%  \\
  \cline{2-7}
  Natural &  BD &   23.77\% &   13.20\% &   12.37\% &   12.34\% &   12.37\%   \\
  \cline{2-7}
    &   BD-F &   15.79\% &   3.78\% &   2.27\% &   1.80\% &   1.51\%  \\
  \cline{1-7}
    &   FM &   9.14\% &   1.64\% &   0.83\% &   0.71\% &   0.62\%  \\
  \cline{2-7}
  CIFAR10  &   FL  &   9.14\% &   1.70\% &   0.93\% &   0.83\% &   0.76\%  \\
  \cline{2-7}
  VGG  &   BM &   73.25\% &   56.16\% &   43.35\% &   34.75\% &   29.28\%  \\
  \cline{2-7}
  100\%  &   BL  &   74.21\% &   57.81\% &   45.44\% &   36.93\% &   30.57\%  \\
  \cline{2-7}
  Natural &  BD &   16.05\% &   9.83\% &   9.62\% &   9.38\% &   9.39\%   \\
  \cline{2-7}
    &   BD-F &   11.13\% &   3.03\% &   1.91\% &   1.57\% &   1.33\%  \\
  \cline{1-7}
   &   FM  &   3.53\% &   0.24\% &   0.13\% &   0.08\% &   0.07\% \\
  \cline{2-7}
  CIFAR100  &   FL  &   4.01\% &   0.31\% &   0.15\% &   0.09\% &   0.07\%  \\
  \cline{2-7}
  ResNet-18  &   BM &   52.02\% &   34.72\% &   23.43\% &   16.88\% &   13.20\%  \\
  \cline{2-7}
  100\%  &   BL  &   53.22\% &   35.69\% &   24.87\% &   17.91\% &   14.40\%  \\
  \cline{2-7}
  Natural &  BD &   9.31\% &   1.44\% &   1.17\% &   1.13\% &   1.17\%   \\
  \cline{2-7}
    &   BD-F &   5.09\% &   0.54\% &   0.34\% &   0.26\% &   0.20\%  \\
  \cline{1-7}
   &   FM &   3.36\% &   0.33\% &   0.19\% &   0.12\% &   0.11\%  \\
  \cline{2-7}
  CIFAR100  &   FL  &   3.41\% &   0.34\% &   0.19\% &   0.14\% &   0.11\%  \\
  \cline{2-7}
  VGG  &   BM &   50.66\% &   33.25\% &   22.13\% &   16.04\% &   12.16\%  \\
  \cline{2-7}
  100\%  &   BL  &   52.02\% &   35.26\% &   23.97\% &   17.46\% &   13.71\%  \\
  \cline{2-7}
  Natural &  BD &   6.75\% &   1.14\% &   0.94\% &   0.86\% &   0.83\%   \\
  \cline{2-7}
    &   BD-F &   4.44\% &   0.53\% &   0.21\% &   0.16\% &   0.16\%  \\
\bottomrule
\end{tabular}
\newline\newline 
\end{adjustbox}
\end{center}
\end{table*}

%% file: mt-tab-sd.tex
\begin{table*}
\begin{center}
\caption{Adaptive metamorphic testing of forward fitness scores and BD scores from~\citet{diversity-boosting}: the first column specifies the data, model, pseudo-accuracy, and underlying data transformations. The second column specifies which metric is used for testing. The rest of the columns are the pseudo-accuracy of tests prioritized by each metric during each iteration in adaptive testing and the accuracy is relative to pseudo-labels rather than ground-truth labels.}\label{tab:mt-sd}
\begin{adjustbox}{width=\linewidth,center}
\begin{tabular}{C{2.8cm}|L{1.2cm}|C{1.4cm}C{1.4cm}C{1.4cm}C{1.4cm}C{1.4cm}}
\toprule
\multirow{3}{*}{\shortstack[c]{\bf Data Model \\ \bf Pseudo-accuracy \\ \bf Transformations}} & \multirow{3}{*}{\bf Fitness} & \multirow{3}{*}{\bf Iter 1} & \multirow{3}{*}{\bf Iter 2} & \multirow{3}{*}{\bf Iter 3} & \multirow{3}{*}{\bf Iter 4} & \multirow{3}{*}{\bf Iter 5} \\ 
& & & & & & \\
& & & & & & \\ \hline \hline
  &   FM &   98.09\% &   95.87\% &   93.50\% &   90.45\% &   86.91\%  \\
\cline{2-7}
  MNIST &   FL  &   98.14\% &   96.03\% &   93.61\% &   90.74\% &   87.35\%   \\
\cline{2-7}
  LeNet-1 &   BM &   70.21\% &   15.51\% &   2.37\% &   0.20\% &   0.04\%  \\
\cline{2-7}
  100\% &   BL  &   33.03\% &   0.65\% &   0.00\% &   0.00\% &   0.00\%   \\
\cline{2-7}
  SD &  BD &   99.59\% &   99.41\% &   99.22\% &   98.81\% &   98.50\%   \\
\cline{2-7}
    &   BD-F &   99.58\% &   99.28\% &   98.99\% &   98.68\% &   98.14\%  \\
\cline{1-7}
  &   FM &   98.94\% &   97.72\% &   96.32\% &   94.46\% &   91.44\%  \\
\cline{2-7}
  MNIST &   FL  &   98.96\% &   97.75\% &   96.39\% &   94.60\% &   91.66\%   \\
\cline{2-7}
  LeNet-5 &   BM &   69.05\% &   13.69\% &   1.41\% &   0.04\% &   0.00\%  \\
\cline{2-7}
  100\% &   BL  &   42.05\% &   1.39\% &   0.01\% &   0.00\% &   0.00\%   \\
\cline{2-7}
  SD &  BD &   99.75\% &   99.58\% &   99.51\% &   99.43\% &   99.26\%   \\
\cline{2-7}
    &   BD-F &   99.69\% &   99.36\% &   99.03\% &   98.42\% &   97.77\%  \\
\cline{1-7}
   &   FM &   99.32\% &   98.69\% &   98.09\% &   97.47\% &   96.79\%  \\
\cline{2-7}
  SVHN &   FL  &   99.34\% &   98.72\% &   98.16\% &   97.57\% &   96.92\%  \\
\cline{2-7}
  ResNet-9 &  BM &   90.40\% &   78.22\% &   64.22\% &   50.72\% &   39.14\%   \\
\cline{2-7}
  100\% &   BL &   87.32\% &   71.41\% &   54.93\% &   41.11\% &   29.95\%  \\
\cline{2-7}
    SD &  BD &   99.91\% &   99.84\% &   99.78\% &   99.75\% &   99.69\%   \\
\cline{2-7}
    &   BD-F &   99.82\% &   99.71\% &   99.56\% &   99.47\% &   98.35\%  \\
\cline{1-7}
   &   FM &   99.33\% &   98.50\% &   97.89\% &   97.21\% &   96.55\%  \\
\cline{2-7}
  SVHN &   FL  &   99.35\% &   98.51\% &   97.91\% &   97.25\% &   96.61\%  \\
\cline{2-7}
  VGG &  BM &  86.33\% &   73.89\% &   63.21\% &   53.46\%  &   44.40\% \\
\cline{2-7}
  100\% &   BL &   85.79\% &   72.11\% &   60.05\% &   49.18\% &   39.51\%  \\
\cline{2-7}
    SD &  BD &   99.80\% &   99.61\% &   99.41\% &   99.17\% &   98.94\%   \\
\cline{2-7}
    &   BD-F &   99.70\% &   99.34\% &   99.07\% &   98.71\% &   98.44\%  \\
\cline{1-7}
   &   FM &   98.05\% &   96.21\% &   94.17\% &   92.56\% &   90.94\%  \\
\cline{2-7}
  CIFAR10 &   FL  &   98.14\% &   96.30\% &   94.22\% &   92.72\% &   91.11\%  \\
\cline{2-7}
  ResNet-18 &   BM &   68.20\% &   42.29\% &   26.16\% &   15.67\% &   9.58\%  \\
\cline{2-7}
  100\% &   BL  &   65.08\% &   37.18\% &   20.81\% &   10.94\% &   6.08\%  \\
\cline{2-7}
  SD &  BD &   99.57\% &   99.31\% &   99.04\% &   98.80\% &   98.46\%   \\
\cline{2-7}
    &   BD-F &   99.25\% &   98.76\% &   98.15\% &   97.59\% &   97.18\%  \\
\bottomrule
\end{tabular}
\newline\newline 
\end{adjustbox}
\end{center}
\end{table*}

\begin{table*}
\begin{center}
\caption*{Table (Continued)}
\begin{adjustbox}{width=\linewidth,center}
\begin{tabular}{C{2.8cm}|L{1.2cm}|C{1.4cm}C{1.4cm}C{1.4cm}C{1.4cm}C{1.4cm}}
\cline{1-7}
   &   FM &   98.69\% &   97.00\% &   95.55\% &   94.29\% &   92.71\%  \\
\cline{2-7}
  CIFAR10 &   FL  &   98.71\% &   97.00\% &   95.63\% &   94.33\% &   92.74\%  \\
\cline{2-7}
  VGG &   BM &   71.34\% &   50.70\% &   36.72\% &   26.69\% &   18.88\%  \\
\cline{2-7}
  100\% &   BL  &   70.38\% &   48.72\% &   34.28\% &   24.23\% &   16.55\%  \\
\cline{2-7}
  SD &  BD &   99.53\% &   98.90\% &   98.38\% &   97.93\% &   97.56\%   \\
\cline{2-7}
    &   BD-F &   99.36\% &   98.67\% &   97.94\% &   97.29\% &   96.68\%  \\
\cline{1-7}
   &   FM &   91.76\% &   84.64\% &   78.53\% &   73.55\% &   69.12\%  \\
\cline{2-7}
  CIFAR100 &   FL  &   93.25\% &   86.99\% &   81.70\% &   77.04\% &   72.67\%  \\
\cline{2-7}
  ResNet-18 &   BM &   38.38\% &   14.01\% &   4.48\% &   1.53\% &   0.58\%  \\
\cline{2-7}
  100\% &   BL  &   34.31\% &   11.40\% &   3.54\% &   1.22\% &   0.42\%  \\
\cline{2-7}
  SD &  BD &   98.12\% &   96.74\% &   95.74\% &   94.98\% &   94.19\%   \\
\cline{2-7}
    &   BD-F &   96.88\% &   94.65\% &   92.38\% &   90.30\% &   88.11\%  \\
\cline{1-7}
   &   FM &   93.46\% &  88.32\% &   83.45\% &   79.25\% &   75.41\%  \\
\cline{2-7}
  CIFAR100 &   FL  &   93.76\% &   88.78\% &   84.11\% &   79.94\% &   76.08\%  \\
\cline{2-7}
  VGG &   BM &   43.49\% &   21.97\% &   11.42\% &   5.74\% &   3.02\%  \\
\cline{2-7}
  100\% &   BL  &   38.36\% &   17.98\% &   8.43\% &   3.86\% &   2.06\%  \\
\cline{2-7}
  SD &  BD &   98.19\% &   97.25\% &   96.13\% &   95.26\% &   94.45\%   \\
\cline{2-7}
    &   BD-F &   97.24\% &   95.07\% &   92.98\% &   90.91\% &   89.10\%  \\
\bottomrule
\end{tabular}
\newline\newline 
\end{adjustbox}
\end{center}
\end{table*}

%% file: mt-tab-mix.tex
\begin{table*}
\begin{center}
\caption{Adaptive metamorphic testing of forward fitness scores and BD scores from~\citet{diversity-boosting}: the first column specifies the data, model, pseudo-accuracy, and underlying data transformations. The second column specifies which metric is used for testing. The rest of the columns are the pseudo-accuracy of tests prioritized by each metric during each iteration in adaptive testing and the accuracy is relative to pseudo-labels rather than ground-truth labels.}\label{tab:mt-mix}
\begin{adjustbox}{width=\linewidth,center}
\begin{tabular}{C{2.8cm}|L{1.2cm}|C{1.4cm}C{1.4cm}C{1.4cm}C{1.4cm}C{1.4cm}}
\toprule
\multirow{4}{*}{\shortstack[c]{\bf Data \\ \bf Model \\ \bf Pseudo-accuracy \\ \bf Transformations}} & \multirow{4}{*}{\bf Fitness} & \multirow{4}{*}{\bf Iter 1} & \multirow{4}{*}{\bf Iter 2} & \multirow{4}{*}{\bf Iter 3} & \multirow{4}{*}{\bf Iter 4} & \multirow{4}{*}{\bf Iter 5} \\ 
& & & & & & \\
& & & & & & \\
& & & & & & \\ \hline \hline
 &  FM &  6.28\% &  0.23\% &  0.06\% &  0.00\% &  0.00\%  \\
\cline{2-7}
  &  FL  &  6.39\% &  0.25\% &  0.06\% &  0.00\% &  0.00\%   \\
\cline{2-7}
 MNIST &  BM &  29.37\% &  11.12\% &  5.70\% &  2.43\% &  1.51\%  \\
\cline{2-7}
 LeNet-1 &  BL  &  42.11\% &  18.20\% &  10.31\% &  5.86\% &  3.63\%   \\
\cline{2-7}
 100\% &  MM &  6.27\% &  0.17\% &  0.01\% &  0.00\% &  0.00\%  \\
\cline{2-7}
 Mixed &  ML  &  6.33\% &  0.09\% &  0.00\% &  0.00\% &  0.00\%   \\
\cline{2-7}
  & BD &  20.58\% &  15.85\% &  14.35\% &  13.79\% &  13.20\%   \\
\cline{2-7}
   &  BD-F &  16.21\% &  7.87\% &  4.86\% &  3.83\% &  2.98\%  \\
\cline{1-7}
 &  FM &  4.39\% &  0.04\% &  0.00\% &  0.00\% &  0.00\%  \\
\cline{2-7}
  &  FL  &  4.41\% &  0.04\% &  0.00\% &  0.00\% &  0.00\%   \\
\cline{2-7}
 MNIST &  BM &  31.55\% &  9.91\% &  4.50\% &  2.08\% &  1.20\%  \\
\cline{2-7}
 LeNet-5 &  BL  &  40.98\% &  15.41\% &  6.71\% &  3.20\% &  1.62\%   \\
\cline{2-7}
 100\% &  MM &  4.37\% &  0.04\% &  0.00\% &  0.00\% &  0.00\%  \\
\cline{2-7}
 Mixed &  ML  &  4.37\% &  0.03\% &  0.00\% &  0.00\% &  0.00\%   \\
\cline{2-7}
  & BD &  23.07\% &  13.01\% &  11.13\% &  10.85\% &  10.58\%   \\
\cline{2-7}
   &  BD-F &  13.27\% &  5.42\% &  3.90\% &  3.13\% &  2.77\%  \\\cline{1-7}
 &  FM &  13.86\% &  1.41\% &  0.30\% &  0.05\% &  0.02\%  \\
\cline{2-7}
  &  FL  &  14.06\% &  1.67\% &  0.42\% &  0.05\% &  0.02\%   \\
\cline{2-7}
 SVHN &  BM &  78.68\% &  54.53\% &  35.91\% &  23.67\% &  15.98\%  \\
\cline{2-7}
 ResNet-9 &  BL  &  76.68\% &  49.62\% &  29.66\% &  17.04\% &  10.16\%   \\
\cline{2-7}
 100\% &  MM &  13.35\% &  0.69\% &  0.05\% &  0.00\% &  0.00\%  \\
\cline{2-7}
 Mixed &  ML  &  13.21\% &  0.65\% &  0.04\% &  0.01\% &  0.00\%   \\
\cline{2-7}
  & BD &  32.21\% &  14.87\% &  11.07\% &  10.49\% &  10.54\%   \\
\cline{2-7}
   &  BD-F &  16.96\% &  3.33\% &  1.51\% &  0.85\% &  0.63\%  \\\cline{1-7}
 &  FM &  15.00\% &  3.76\% &  2.47\% &  2.06\% &  1.82\%  \\
\cline{2-7}
  &  FL  &  15.07\% &  4.31\% &  3.04\% &  2.56\% &  2.31\%   \\
\cline{2-7}
 SVNH &  BM &  72.28\% &  48.64\% &  32.46\% &  21.00\% &  13.89\%  \\
\cline{2-7}
 VGG &  BL  &  73.89\% &  50.65\% &  32.99\% &  20.39\% &  12.55\%   \\
\cline{2-7}
 100\% &  MM &  13.96\% &  0.40\% &  0.01\% &  0.01\% &  0.00\% \\
\cline{2-7}
 Mixed &  ML  &  13.90\% &  0.48\% &  0.03\% &  0.01\% &  0.00\%   \\
\cline{2-7}
  & BD &  26.36\% &  19.38\% &  19.69\% &  19.81\% &  19.84\%   \\
\cline{2-7}
   &  BD-F &  16.77\% &  5.81\% &  4.29\% &  3.67\% &  3.30\%  \\
\bottomrule
\end{tabular}
\newline\newline 
\end{adjustbox}
\end{center}
\end{table*}

\begin{table*}
\begin{center}
\caption*{Table (Continued)}
\begin{adjustbox}{width=\linewidth,center}
\begin{tabular}{C{2.8cm}|L{1.2cm}|C{1.4cm}C{1.4cm}C{1.4cm}C{1.4cm}C{1.4cm}}
\cline{1-7}
 &  FM &  8.93\% &  1.06\% &  0.26\% &  0.15\% &  0.10\%  \\
\cline{2-7}
  &  FL  &  9.06\% &  1.14\% &  0.34\% &  0.19\% &  0.12\%   \\
\cline{2-7}
 CIFAR10 &  BM &  58.04\% &  32.13\% &  18.93\% &  12.58\% &  8.82\%  \\
\cline{2-7}
 ResNet-18 &  BL  &  53.43\% &  24.31\% &  10.65\% &  4.86\% &  2.14\%   \\
\cline{2-7}
 100\% &  MM &  7.15\% &  0.36\% &  0.05\% &  0.01\% &  0.00\%  \\
\cline{2-7}
 Mixed &  ML  &  7.03\% &  0.33\% &  0.04\% &  0.00\% &  0.00\%   \\
\cline{2-7}
  & BD &  20.66\% &  13.00\% &  12.48\% &  12.76\% &  12.64\%   \\
\cline{2-7}
   &  BD-F &  13.08\% &  3.42\% &  2.22\% &  1.68\% &  1.36\%  \\
\cline{1-7}
 &  FM &  7.04\% &  1.64\% &  1.07\% &  0.80\% &  0.60\%  \\
\cline{2-7}
  &  FL  &  7.05\% &  1.68\% &  1.16\% &  0.96\% &  0.88\%   \\
\cline{2-7}
 CIFAR10 &  BM &  62.62\% &  40.87\% &  26.85\% &  18.30\% &  12.16\%  \\
\cline{2-7}
 VGG &  BL  &  61.87\% &  38.33\% &  22.66\% &  13.33\% &  7.64\%   \\
\cline{2-7}
 100\% &  MM &  5.29\% &  0.42\% &  0.10\% &  0.00\% &  0.00\%  \\
\cline{2-7}
 Mixed &  ML  &  5.22\% &  0.43\% &  0.08\% &  0.01\% &  0.00\%   \\
\cline{2-7}
  & BD &  15.54\% &  9.71\% &  9.60\% &  9.61\% &  9.63\%   \\
\cline{2-7}
   &  BD-F &  9.10\% &  3.08\% &  2.21\% &  1.73\% &  1.46\%  \\\cline{1-7}
 &  FM &  2.60\% &  0.21\% &  0.07\% &  0.03\% &  0.01\%  \\
\cline{2-7}
  &  FL  &  2.94\% &  0.33\% &  0.12\% &  0.04\% &  0.03\%   \\
\cline{2-7}
 CIFAR100 &  BM &  28.65\% &  12.03\% &  8.27\% &  5.77\% &  3.53\%  \\
\cline{2-7}
 ResNet-18 &  BL  &  23.28\% &  4.82\% &  1.48\% &  0.71\% &  0.52\%   \\
\cline{2-7}
 100\% &  MM &  1.80\% &  0.01\% &  0.00\% &  0.00\% &  0.00\%  \\
\cline{2-7}
 Mixed &  ML  &  1.91\% &  0.02\% &  0.00\% &  0.00\% &  0.00\%   \\
\cline{2-7}
  & BD &  6.88\% &  1.49\% &  1.24\% &  1.18\% &  1.16\%   \\
\cline{2-7}
   &  BD-F &  3.97\% &  0.69\% &  0.42\% &  0.27\% &  0.26\%  \\
\cline{1-7}
 &  FM &  2.31\% &  0.27\% &  0.09\% &  0.05\% &  0.03\%  \\
\cline{2-7}
  &  FL  &  2.33\% &  0.28\% &  0.11\% &  0.05\% &  0.04\%   \\
\cline{2-7}
 CIFAR100 &  BM &  36.10\% &  17.51\% &  9.48\% &  6.57\% &  4.39\%  \\
\cline{2-7}
 VGG &  BL  &  29.46\% &  11.30\% &  4.33\% &  2.36\% &  1.46\%   \\
\cline{2-7}
 100\% &  MM &  1.82\% &  0.08\% &  0.01\% &  0.01\% &  0.00\%  \\
\cline{2-7}
 Mixed &  ML  &  1.78\% &  0.05\% &  0.01\% &  0.01\% &  0.01\%   \\
\cline{2-7}
  & BD &  5.28\% &  1.01\% &  0.89\% &  0.89\% &  0.87\%   \\
\cline{2-7}
   &  BD-F &  3.03\% &  0.42\% &  0.17\% &  0.15\% &  0.12\%  \\
\bottomrule
\end{tabular}
\newline\newline
\end{adjustbox}

\end{center}
\end{table*}

%% file: tab-iter3.tex
\begin{table*}
\begin{center}
\caption{Adaptive testing of forward fitness scores and BD scores from~\citet{diversity-boosting} for 3 iterations: the first column specifies the data, model, clean accuracy, and underlying data transformations. The second column specifies which metric is used for testing. The rest of the columns are the accuracy of tests prioritized by each metric during each iteration in adaptive testing.}\label{tab:iter3}
\begin{tabular}{p{6cm}p{2.9cm}p{1cm}p{1.2cm}p{1.3cm}}
\toprule
\centering Data, Model, Accuracy, Transformations & \hfil Fitness &\hfil Iter 1 & \hfil Iter 2 & \hfil Iter 3   \\
\midrule
\hfil MNIST, & \hfil FM & \hfil 4.67\% & \hfil 0.01\% & \hfil 0.00\%  \\
\cmidrule{2-5}
\hfil LeNet-5, & \hfil FL  & \hfil 4.73\% & \hfil 0.01\% & \hfil 0.00\%   \\
\cmidrule{2-5}
\centering 99.01\%, &\hfil BD & \hfil 21.50\% & \hfil 10.90\% & \hfil 8.88\%   \\
\cmidrule{2-5}
\centering  Benign & \hfil BD-F & \hfil 13.51\% & \hfil 3.69\% & \hfil 2.09\% \\
\midrule
\hfil SVHN, & \hfil FM & \hfil 9.27\% & \hfil 0.81\% & \hfil 0.23\%  \\
\cmidrule{2-5}
\hfil ResNet-9, & \hfil FL  & \hfil 9.47\% & \hfil 0.93\% & \hfil 0.32\%  \\
\cmidrule{2-5}
\centering 95.90\%, &\hfil BD & \hfil 28.02\% & \hfil 13.10\% & \hfil 10.79\%   \\
\cmidrule{2-5}
\centering  Benign & \hfil BD-F & \hfil 12.41\% & \hfil 2.68\% & \hfil 1.49\%  \\
\midrule
\hfil CIFAR10, & \hfil FM & \hfil 5.44\% & \hfil 0.84\% & \hfil 0.47\%   \\
\cmidrule{2-5}
\hfil VGG, & \hfil FL  & \hfil 5.50\% & \hfil 0.88\% & \hfil 0.53\% \\
\cmidrule{2-5}
\centering 91.48\%, &\hfil BD & \hfil 13.97\% & \hfil 9.68\% & \hfil 9.42\%  \\
\cmidrule{2-5}
\centering  Benign & \hfil BD-F & \hfil 8.49\% & \hfil 2.35\% & \hfil 1.64\%  \\
\midrule
\hfil CIFAR100, & \hfil FM & \hfil 1.82\% & \hfil 0.15\% & \hfil 0.03\%  \\
\cmidrule{2-5}
\hfil ResNet-18, & \hfil FL  & \hfil 2.03\% & \hfil 0.20\% & \hfil 0.05\%  \\
\cmidrule{2-5}
\centering 70.45\%, &\hfil BD & \hfil 5.08\% & \hfil 1.34\% & \hfil 1.13\%   \\
\cmidrule{2-5}
\centering  Benign & \hfil BD-F & \hfil 2.66\% & \hfil 0.41\% & \hfil 0.22\%  \\
\bottomrule
\end{tabular}
\newline\newline 
\end{center}
\end{table*}

%% file: tab-iter8.tex
\begin{table*}
\begin{center}
\caption{Adaptive testing of forward fitness scores and BD scores from~\citet{diversity-boosting} for 8 iterations: the first column specifies the data, model, clean accuracy, and underlying data transformations. The second column specifies which metric is used for testing. The rest of the columns are the accuracy of tests prioritized by each metric during each iteration in adaptive testing.}\label{tab:it8}
\begin{tabular}{p{3.2cm}p{1.6cm}p{1cm}p{1.1cm}p{1.2cm}p{1.1cm}p{1.1cm}p{1.1cm}p{1.1cm}p{1.1cm}}
\toprule
\centering Data, Model, Accuracy, Transformations & \hfil Fitness &\hfil Iter 1 & \hfil Iter 2 & \hfil Iter 3  & \hfil Iter 4 & \hfil Iter 5 & \hfil Iter 6  & \hfil Iter 7 & \hfil Iter 8  \\
\midrule
\hfil MNIST, & \hfil FM & \hfil 8.36\% & \hfil 0.15\% & \hfil 0.0\% & \hfil 0.0\% & \hfil 0.0\% & \hfil 0.0\% & \hfil 0.0\% & \hfil 0.0\% \\
\cmidrule{2-10}
\hfil LeNet-5, & \hfil FL  & \hfil 8.38\% & \hfil 0.16\% & \hfil 0.0\% & \hfil 0.0\% & \hfil 0.0\% & \hfil 0.0\% & \hfil 0.0\% & \hfil 0.0\% \\
\cmidrule{2-10}
\centering 99.01\%, &\hfil BD & \hfil 31.11\% & \hfil 12.80\% & \hfil 10.53\% & \hfil 9.66\% & \hfil 8.92\% & \hfil 8.58\% & \hfil 8.53\% & \hfil 8.56\%  \\
\cmidrule{2-10}
\centering  Benign & \hfil BD-F & \hfil 14.28\% & \hfil 2.87\% & \hfil 2.04\% & \hfil 1.78\% & \hfil 1.43\% & \hfil 1.32\% & \hfil 1.23\% & \hfil 1.15\%  \\
\midrule
\hfil SVHN, & \hfil FM & \hfil 17.16\% & \hfil 1.86\% & \hfil 0.54\% & \hfil 0.35\% & \hfil 0.30\% & \hfil 0.28\% & \hfil 0.27\% & \hfil 0.25\%  \\
\cmidrule{2-10}
\hfil ResNet-9, & \hfil FL  & \hfil 17.35\% & \hfil 2.05\% & \hfil 0.66\% & \hfil 0.43\% & \hfil 0.35\% & \hfil 0.32\% & \hfil 0.29\% & \hfil 0.29\%  \\
\cmidrule{2-10}
\centering 95.90\%, &\hfil BD & \hfil 36.73\% & \hfil 14.34\% & \hfil 10.53\% & \hfil 9.92\% & \hfil 9.83\% & \hfil 9.80\% & \hfil 9.85\% & \hfil 9.84\%   \\
\cmidrule{2-10}
\centering  Benign & \hfil BD-F & \hfil 20.76\% & \hfil 4.22\% & \hfil 2.00\% & \hfil 1.52\% & \hfil 1.29\% & \hfil 1.19\% & \hfil 1.09\% & \hfil 1.03\%  \\
\midrule
\hfil CIFAR10, & \hfil FM & \hfil 9.41\% & \hfil 1.33\% & \hfil 0.74\% & \hfil 0.59\% & \hfil 0.51\% & \hfil 0.46\% & \hfil 0.41\% & \hfil 0.35\% \\
\cmidrule{2-10}
\hfil VGG, & \hfil FL  & \hfil 9.44\% & \hfil 1.46\% & \hfil 0.86\% & \hfil 0.73\% & \hfil 0.64\% & \hfil 0.58\% & \hfil 0.52\% & \hfil 0.45\%   \\
\cmidrule{2-10}
\centering 91.48\%, &\hfil BD & \hfil 17.26\% & \hfil 9.99\% & \hfil 9.72\% & \hfil 9.56\% & \hfil 9.49\% & \hfil 9.46\% & \hfil 9.43\% & \hfil 9.44\%   \\
\cmidrule{2-10}
\centering  Benign & \hfil BD-F & \hfil 13.25\% & \hfil 3.40\% & \hfil 2.45\% & \hfil 2.02\% & \hfil 1.88\% & \hfil 1.67\% & \hfil 1.50\% & \hfil 1.34\%  \\
\midrule
\hfil CIFAR100, & \hfil FM & \hfil 3.69\% & \hfil 0.31\% & \hfil 0.06\% &\hfil 0.04\% & \hfil 0.02\% & \hfil 0.01\% & \hfil 0.00\% & \hfil 0.00\%  \\
\cmidrule{2-10}
\hfil ResNet-18, & \hfil FL  & \hfil 4.10\% & \hfil 0.35\% & \hfil 0.11\% &\hfil 0.07\% & \hfil 0.03\% & \hfil 0.02\% & \hfil 0.01\% & \hfil 0.01\%  \\
\cmidrule{2-10}
\centering 70.45\%, &\hfil BD & \hfil 9.22\% & \hfil 1.26\% & \hfil 1.07\% & \hfil 1.04\% & \hfil 1.05\%  & \hfil 1.02\% & \hfil 1.01\% & \hfil 0.98\%  \\
\cmidrule{2-10}
\centering  Benign & \hfil BD-F & \hfil 5.68\% & \hfil 0.64\% & \hfil 0.44\% & \hfil 0.34\% & \hfil 0.26\% & \hfil 0.27\% & \hfil 0.25\% & \hfil 0.25\% \\
\bottomrule
\end{tabular}
\newline\newline 
\end{center}
\end{table*}

%% file: paper.bbl

\begin{thebibliography}{39}


\ifx \showCODEN    \undefined \def \showCODEN     #1{\unskip}     \fi
\ifx \showDOI      \undefined \def \showDOI       #1{#1}\fi
\ifx \showISBNx    \undefined \def \showISBNx     #1{\unskip}     \fi
\ifx \showISBNxiii \undefined \def \showISBNxiii  #1{\unskip}     \fi
\ifx \showISSN     \undefined \def \showISSN      #1{\unskip}     \fi
\ifx \showLCCN     \undefined \def \showLCCN      #1{\unskip}     \fi
\ifx \shownote     \undefined \def \shownote      #1{#1}          \fi
\ifx \showarticletitle \undefined \def \showarticletitle #1{#1}   \fi
\ifx \showURL      \undefined \def \showURL       {\relax}        \fi
\providecommand\bibfield[2]{#2}
\providecommand\bibinfo[2]{#2}
\providecommand\natexlab[1]{#1}
\providecommand\showeprint[2][]{arXiv:#2}

\bibitem[Adebayo et~al\mbox{.}(2020)]%
        {debug}
\bibfield{author}{\bibinfo{person}{Julius Adebayo}, \bibinfo{person}{Michael Muelly}, \bibinfo{person}{Ilaria Liccardi}, {and} \bibinfo{person}{Been Kim}.} \bibinfo{year}{2020}\natexlab{}.
\newblock \showarticletitle{Debugging Tests for Model Explanations}. In \bibinfo{booktitle}{\emph{Proceedings of the 34th International Conference on Neural Information Processing Systems}} (Vancouver, BC, Canada) \emph{(\bibinfo{series}{NIPS'20})}. \bibinfo{publisher}{Curran Associates Inc.}, \bibinfo{address}{Red Hook, NY, USA}, Article \bibinfo{articleno}{60}, \bibinfo{numpages}{13}~pages.
\newblock
\showISBNx{9781713829546}


\bibitem[B\"{o}hme et~al\mbox{.}(2017)]%
        {aflgo}
\bibfield{author}{\bibinfo{person}{Marcel B\"{o}hme}, \bibinfo{person}{Van-Thuan Pham}, \bibinfo{person}{Manh-Dung Nguyen}, {and} \bibinfo{person}{Abhik Roychoudhury}.} \bibinfo{year}{2017}\natexlab{}.
\newblock \showarticletitle{Directed Greybox Fuzzing}. In \bibinfo{booktitle}{\emph{Proceedings of the 2017 ACM SIGSAC Conference on Computer and Communications Security}} (Dallas, Texas, USA) \emph{(\bibinfo{series}{CCS '17})}. \bibinfo{publisher}{Association for Computing Machinery}, \bibinfo{address}{New York, NY, USA}, \bibinfo{pages}{2329–2344}.
\newblock
\showISBNx{9781450349468}
\urldef\tempurl%
\url{https://doi.org/10.1145/3133956.3134020}
\showDOI{\tempurl}


\bibitem[Bolukbasi et~al\mbox{.}(2016)]%
        {nn-bias}
\bibfield{author}{\bibinfo{person}{Tolga Bolukbasi}, \bibinfo{person}{Kai-Wei Chang}, \bibinfo{person}{James Zou}, \bibinfo{person}{Venkatesh Saligrama}, {and} \bibinfo{person}{Adam Kalai}.} \bibinfo{year}{2016}\natexlab{}.
\newblock \showarticletitle{Man is to Computer Programmer as Woman is to Homemaker? Debiasing Word Embeddings}. In \bibinfo{booktitle}{\emph{Proceedings of the 30th International Conference on Neural Information Processing Systems}} (Barcelona, Spain) \emph{(\bibinfo{series}{NIPS'16})}. \bibinfo{publisher}{Curran Associates Inc.}, \bibinfo{address}{Red Hook, NY, USA}, \bibinfo{pages}{4356–4364}.
\newblock
\showISBNx{9781510838819}


\bibitem[Chen et~al\mbox{.}(2021)]%
        {metamorphic}
\bibfield{author}{\bibinfo{person}{Songqiang Chen}, \bibinfo{person}{Shuo Jin}, {and} \bibinfo{person}{Xiaoyuan Xie}.} \bibinfo{year}{2021}\natexlab{}.
\newblock \showarticletitle{Validation on Machine Reading Comprehension Software without Annotated Labels: A Property-Based Method}. In \bibinfo{booktitle}{\emph{Proceedings of the 29th ACM Joint Meeting on European Software Engineering Conference and Symposium on the Foundations of Software Engineering}} (Athens, Greece) \emph{(\bibinfo{series}{ESEC/FSE 2021})}. \bibinfo{publisher}{Association for Computing Machinery}, \bibinfo{address}{New York, NY, USA}, \bibinfo{pages}{590–602}.
\newblock
\showISBNx{9781450385626}
\urldef\tempurl%
\url{https://doi.org/10.1145/3468264.3468569}
\showDOI{\tempurl}


\bibitem[Dwork et~al\mbox{.}(2012)]%
        {fairness}
\bibfield{author}{\bibinfo{person}{Cynthia Dwork}, \bibinfo{person}{Moritz Hardt}, \bibinfo{person}{Toniann Pitassi}, \bibinfo{person}{Omer Reingold}, {and} \bibinfo{person}{Richard Zemel}.} \bibinfo{year}{2012}\natexlab{}.
\newblock \showarticletitle{Fairness through awareness}. In \bibinfo{booktitle}{\emph{Proceedings of the 3rd innovations in theoretical computer science conference}}. \bibinfo{pages}{214--226}.
\newblock


\bibitem[Gao et~al\mbox{.}(2022)]%
        {ATS}
\bibfield{author}{\bibinfo{person}{Xinyu Gao}, \bibinfo{person}{Yang Feng}, \bibinfo{person}{Yining Yin}, \bibinfo{person}{Zixi Liu}, \bibinfo{person}{Zhenyu Chen}, {and} \bibinfo{person}{Baowen Xu}.} \bibinfo{year}{2022}\natexlab{}.
\newblock \showarticletitle{Adaptive test selection for deep neural networks}. In \bibinfo{booktitle}{\emph{2022 IEEE/ACM 44th International Conference on Software Engineering (ICSE)}}. IEEE, \bibinfo{pages}{73--85}.
\newblock


\bibitem[Godefroid(2020)]%
        {fuzzing-cacm}
\bibfield{author}{\bibinfo{person}{Patrice Godefroid}.} \bibinfo{year}{2020}\natexlab{}.
\newblock \showarticletitle{Fuzzing: Hack, Art, and Science}.
\newblock \bibinfo{journal}{\emph{Commun. ACM}} \bibinfo{volume}{63}, \bibinfo{number}{2} (\bibinfo{date}{jan} \bibinfo{year}{2020}), \bibinfo{pages}{70–76}.
\newblock
\showISSN{0001-0782}
\urldef\tempurl%
\url{https://doi.org/10.1145/3363824}
\showDOI{\tempurl}


\bibitem[Goodfellow et~al\mbox{.}(2016)]%
        {deep-learning}
\bibfield{author}{\bibinfo{person}{Ian Goodfellow}, \bibinfo{person}{Yoshua Bengio}, {and} \bibinfo{person}{Aaron Courville}.} \bibinfo{year}{2016}\natexlab{}.
\newblock \bibinfo{booktitle}{\emph{Deep Learning}}.
\newblock \bibinfo{publisher}{MIT Press}.
\newblock
\newblock
\shownote{\url{http://www.deeplearningbook.org}}.


\bibitem[Goodfellow et~al\mbox{.}(2014)]%
        {gan}
\bibfield{author}{\bibinfo{person}{Ian Goodfellow}, \bibinfo{person}{Jean Pouget-Abadie}, \bibinfo{person}{Mehdi Mirza}, \bibinfo{person}{Bing Xu}, \bibinfo{person}{David Warde-Farley}, \bibinfo{person}{Sherjil Ozair}, \bibinfo{person}{Aaron Courville}, {and} \bibinfo{person}{Yoshua Bengio}.} \bibinfo{year}{2014}\natexlab{}.
\newblock \showarticletitle{Generative Adversarial Nets}. In \bibinfo{booktitle}{\emph{Advances in Neural Information Processing Systems}}, \bibfield{editor}{\bibinfo{person}{Z.~Ghahramani}, \bibinfo{person}{M.~Welling}, \bibinfo{person}{C.~Cortes}, \bibinfo{person}{N.~Lawrence}, {and} \bibinfo{person}{K.Q. Weinberger}} (Eds.), Vol.~\bibinfo{volume}{27}. \bibinfo{publisher}{Curran Associates, Inc.}
\newblock
\urldef\tempurl%
\url{https://proceedings.neurips.cc/paper_files/paper/2014/file/5ca3e9b122f61f8f06494c97b1afccf3-Paper.pdf}
\showURL{%
\tempurl}


\bibitem[Goodfellow et~al\mbox{.}(2015)]%
        {goodfellow2015explaining}
\bibfield{author}{\bibinfo{person}{Ian~J. Goodfellow}, \bibinfo{person}{Jonathon Shlens}, {and} \bibinfo{person}{Christian Szegedy}.} \bibinfo{year}{2015}\natexlab{}.
\newblock \showarticletitle{Explaining and Harnessing Adversarial Examples}. In \bibinfo{booktitle}{\emph{3rd International Conference on Learning Representations, {ICLR} 2015, San Diego, CA, USA, May 7-9, 2015, Conference Track Proceedings}}, \bibfield{editor}{\bibinfo{person}{Yoshua Bengio} {and} \bibinfo{person}{Yann LeCun}} (Eds.).
\newblock
\urldef\tempurl%
\url{http://arxiv.org/abs/1412.6572}
\showURL{%
\tempurl}


\bibitem[Harel-Canada et~al\mbox{.}(2020)]%
        {NC-meaningful}
\bibfield{author}{\bibinfo{person}{Fabrice Harel-Canada}, \bibinfo{person}{Lingxiao Wang}, \bibinfo{person}{Muhammad~Ali Gulzar}, \bibinfo{person}{Quanquan Gu}, {and} \bibinfo{person}{Miryung Kim}.} \bibinfo{year}{2020}\natexlab{}.
\newblock \showarticletitle{Is Neuron Coverage a Meaningful Measure for Testing Deep Neural Networks?}. In \bibinfo{booktitle}{\emph{Proceedings of the 28th ACM Joint Meeting on European Software Engineering Conference and Symposium on the Foundations of Software Engineering}} (Virtual Event, USA) \emph{(\bibinfo{series}{ESEC/FSE 2020})}. \bibinfo{publisher}{Association for Computing Machinery}, \bibinfo{address}{New York, NY, USA}, \bibinfo{pages}{851–862}.
\newblock
\showISBNx{9781450370431}
\urldef\tempurl%
\url{https://doi.org/10.1145/3368089.3409754}
\showDOI{\tempurl}


\bibitem[He et~al\mbox{.}(2016)]%
        {resnet}
\bibfield{author}{\bibinfo{person}{Kaiming He}, \bibinfo{person}{Xiangyu Zhang}, \bibinfo{person}{Shaoqing Ren}, {and} \bibinfo{person}{Jian Sun}.} \bibinfo{year}{2016}\natexlab{}.
\newblock \showarticletitle{Deep Residual Learning for Image Recognition}. In \bibinfo{booktitle}{\emph{2016 IEEE Conference on Computer Vision and Pattern Recognition (CVPR)}}. \bibinfo{pages}{770--778}.
\newblock
\urldef\tempurl%
\url{https://doi.org/10.1109/CVPR.2016.90}
\showDOI{\tempurl}


\bibitem[Jeong et~al\mbox{.}(2019)]%
        {Razzer}
\bibfield{author}{\bibinfo{person}{Dae~R. Jeong}, \bibinfo{person}{Kyungtae Kim}, \bibinfo{person}{Basavesh Shivakumar}, \bibinfo{person}{Byoungyoung Lee}, {and} \bibinfo{person}{Insik Shin}.} \bibinfo{year}{2019}\natexlab{}.
\newblock \showarticletitle{Razzer: Finding Kernel Race Bugs through Fuzzing}. In \bibinfo{booktitle}{\emph{2019 IEEE Symposium on Security and Privacy (SP)}}. \bibinfo{pages}{754--768}.
\newblock
\urldef\tempurl%
\url{https://doi.org/10.1109/SP.2019.00017}
\showDOI{\tempurl}


\bibitem[Kang and Lo(2022)]%
        {api-misuse2}
\bibfield{author}{\bibinfo{person}{Hong~Jin Kang} {and} \bibinfo{person}{David Lo}.} \bibinfo{year}{2022}\natexlab{}.
\newblock \showarticletitle{Active Learning of Discriminative Subgraph Patterns for API Misuse Detection}.
\newblock \bibinfo{journal}{\emph{IEEE Transactions on Software Engineering}} \bibinfo{volume}{48}, \bibinfo{number}{8} (\bibinfo{year}{2022}), \bibinfo{pages}{2761--2783}.
\newblock
\urldef\tempurl%
\url{https://doi.org/10.1109/TSE.2021.3069978}
\showDOI{\tempurl}


\bibitem[Krizhevsky et~al\mbox{.}(2009)]%
        {cifar10}
\bibfield{author}{\bibinfo{person}{Alex Krizhevsky}, \bibinfo{person}{Geoffrey Hinton}, {et~al\mbox{.}}} \bibinfo{year}{2009}\natexlab{}.
\newblock \showarticletitle{Learning multiple layers of features from tiny images}.
\newblock  (\bibinfo{year}{2009}).
\newblock


\bibitem[Krizhevsky et~al\mbox{.}(2017)]%
        {Krizhevsky_imagenetclassification}
\bibfield{author}{\bibinfo{person}{Alex Krizhevsky}, \bibinfo{person}{Ilya Sutskever}, {and} \bibinfo{person}{Geoffrey~E. Hinton}.} \bibinfo{year}{2017}\natexlab{}.
\newblock \showarticletitle{ImageNet Classification with Deep Convolutional Neural Networks}.
\newblock \bibinfo{journal}{\emph{Commun. ACM}} \bibinfo{volume}{60}, \bibinfo{number}{6} (\bibinfo{date}{may} \bibinfo{year}{2017}), \bibinfo{pages}{84–90}.
\newblock
\showISSN{0001-0782}
\urldef\tempurl%
\url{https://doi.org/10.1145/3065386}
\showDOI{\tempurl}


\bibitem[LeCun et~al\mbox{.}(1998)]%
        {lenet}
\bibfield{author}{\bibinfo{person}{Yann LeCun}, \bibinfo{person}{L{\'e}on Bottou}, \bibinfo{person}{Yoshua Bengio}, {and} \bibinfo{person}{Patrick Haffner}.} \bibinfo{year}{1998}\natexlab{}.
\newblock \showarticletitle{Gradient-based learning applied to document recognition}.
\newblock \bibinfo{journal}{\emph{Proc. IEEE}} \bibinfo{volume}{86}, \bibinfo{number}{11} (\bibinfo{year}{1998}), \bibinfo{pages}{2278--2324}.
\newblock


\bibitem[LeCun and Cortes(2010)]%
        {mnist}
\bibfield{author}{\bibinfo{person}{Yann LeCun} {and} \bibinfo{person}{Corinna Cortes}.} \bibinfo{year}{2010}\natexlab{}.
\newblock \showarticletitle{{MNIST} handwritten digit database}.
\newblock \bibinfo{howpublished}{http://yann.lecun.com/exdb/mnist/}.
\newblock  (\bibinfo{year}{2010}).
\newblock
\urldef\tempurl%
\url{http://yann.lecun.com/exdb/mnist/}
\showURL{%
\tempurl}


\bibitem[Leino et~al\mbox{.}(2021)]%
        {gloro}
\bibfield{author}{\bibinfo{person}{Klas Leino}, \bibinfo{person}{Zifan Wang}, {and} \bibinfo{person}{Matt Fredrikson}.} \bibinfo{year}{2021}\natexlab{}.
\newblock \showarticletitle{Globally-Robust Neural Networks}. In \bibinfo{booktitle}{\emph{International Conference on Machine Learning (ICML)}}.
\newblock


\bibitem[Ma et~al\mbox{.}(2018)]%
        {deep_gauge}
\bibfield{author}{\bibinfo{person}{Lei Ma}, \bibinfo{person}{Felix Juefei-Xu}, \bibinfo{person}{Fuyuan Zhang}, \bibinfo{person}{Jiyuan Sun}, \bibinfo{person}{Minhui Xue}, \bibinfo{person}{Bo Li}, \bibinfo{person}{Chunyang Chen}, \bibinfo{person}{Ting Su}, \bibinfo{person}{Li Li}, \bibinfo{person}{Yang Liu}, \bibinfo{person}{Jianjun Zhao}, {and} \bibinfo{person}{Yadong Wang}.} \bibinfo{year}{2018}\natexlab{}.
\newblock \showarticletitle{DeepGauge: Multi-Granularity Testing Criteria for Deep Learning Systems}. In \bibinfo{booktitle}{\emph{Proceedings of the 33rd ACM/IEEE International Conference on Automated Software Engineering}} (Montpellier, France) \emph{(\bibinfo{series}{ASE '18})}. \bibinfo{publisher}{Association for Computing Machinery}, \bibinfo{address}{New York, NY, USA}, \bibinfo{pages}{120–131}.
\newblock
\showISBNx{9781450359375}
\urldef\tempurl%
\url{https://doi.org/10.1145/3238147.3238202}
\showDOI{\tempurl}


\bibitem[Madry et~al\mbox{.}(2018)]%
        {pgd}
\bibfield{author}{\bibinfo{person}{Aleksander Madry}, \bibinfo{person}{Aleksandar Makelov}, \bibinfo{person}{Ludwig Schmidt}, \bibinfo{person}{Dimitris Tsipras}, {and} \bibinfo{person}{Adrian Vladu}.} \bibinfo{year}{2018}\natexlab{}.
\newblock \showarticletitle{Towards Deep Learning Models Resistant to Adversarial Attacks}. In \bibinfo{booktitle}{\emph{International Conference on Learning Representations}}.
\newblock


\bibitem[Mikolov et~al\mbox{.}(2013)]%
        {nlp_2013}
\bibfield{author}{\bibinfo{person}{Tomas Mikolov}, \bibinfo{person}{Ilya Sutskever}, \bibinfo{person}{Kai Chen}, \bibinfo{person}{Greg Corrado}, {and} \bibinfo{person}{Jeffrey Dean}.} \bibinfo{year}{2013}\natexlab{}.
\newblock \showarticletitle{Distributed Representations of Words and Phrases and Their Compositionality}. In \bibinfo{booktitle}{\emph{Proceedings of the 26th International Conference on Neural Information Processing Systems - Volume 2}} (Lake Tahoe, Nevada) \emph{(\bibinfo{series}{NIPS'13})}. \bibinfo{publisher}{Curran Associates Inc.}, \bibinfo{address}{Red Hook, NY, USA}, \bibinfo{pages}{3111–3119}.
\newblock


\bibitem[Mueller et~al\mbox{.}(2023)]%
        {cert_box}
\bibfield{author}{\bibinfo{person}{Mark~Niklas Mueller}, \bibinfo{person}{Franziska Eckert}, \bibinfo{person}{Marc Fischer}, {and} \bibinfo{person}{Martin Vechev}.} \bibinfo{year}{2023}\natexlab{}.
\newblock \showarticletitle{Certified Training: Small Boxes are All You Need}. In \bibinfo{booktitle}{\emph{The Eleventh International Conference on Learning Representations}}.
\newblock


\bibitem[Netzer et~al\mbox{.}(2011)]%
        {svhn}
\bibfield{author}{\bibinfo{person}{Yuval Netzer}, \bibinfo{person}{Tao Wang}, \bibinfo{person}{Adam Coates}, \bibinfo{person}{Alessandro Bissacco}, \bibinfo{person}{Bo Wu}, {and} \bibinfo{person}{Andrew~Y. Ng}.} \bibinfo{year}{2011}\natexlab{}.
\newblock \showarticletitle{Reading Digits in Natural Images with Unsupervised Feature Learning}. In \bibinfo{booktitle}{\emph{NIPS Workshop on Deep Learning and Unsupervised Feature Learning 2011}}.
\newblock
\urldef\tempurl%
\url{http://ufldl.stanford.edu/housenumbers/nips2011_housenumbers.pdf}
\showURL{%
\tempurl}


\bibitem[Pavlitskaya et~al\mbox{.}(2022)]%
        {nc-person}
\bibfield{author}{\bibinfo{person}{S. Pavlitskaya}, \bibinfo{person}{S. Yikmis}, {and} \bibinfo{person}{J. Zollner}.} \bibinfo{year}{2022}\natexlab{}.
\newblock \showarticletitle{Is Neuron Coverage Needed to Make Person Detection More Robust?}. In \bibinfo{booktitle}{\emph{2022 IEEE/CVF Conference on Computer Vision and Pattern Recognition Workshops (CVPRW)}}. \bibinfo{publisher}{IEEE Computer Society}, \bibinfo{address}{Los Alamitos, CA, USA}, \bibinfo{pages}{2888--2896}.
\newblock
\urldef\tempurl%
\url{https://doi.org/10.1109/CVPRW56347.2022.00326}
\showDOI{\tempurl}


\bibitem[Pei et~al\mbox{.}(2019)]%
        {deepxplore}
\bibfield{author}{\bibinfo{person}{Kexin Pei}, \bibinfo{person}{Yinzhi Cao}, \bibinfo{person}{Junfeng Yang}, {and} \bibinfo{person}{Suman Jana}.} \bibinfo{year}{2019}\natexlab{}.
\newblock \showarticletitle{DeepXplore: Automated Whitebox Testing of Deep Learning Systems}.
\newblock \bibinfo{journal}{\emph{Commun. ACM}} \bibinfo{volume}{62}, \bibinfo{number}{11} (\bibinfo{date}{oct} \bibinfo{year}{2019}), \bibinfo{pages}{137–145}.
\newblock
\showISSN{0001-0782}
\urldef\tempurl%
\url{https://doi.org/10.1145/3361566}
\showDOI{\tempurl}


\bibitem[Rombach et~al\mbox{.}(2022)]%
        {diffusion}
\bibfield{author}{\bibinfo{person}{Robin Rombach}, \bibinfo{person}{Andreas Blattmann}, \bibinfo{person}{Dominik Lorenz}, \bibinfo{person}{Patrick Esser}, {and} \bibinfo{person}{Bj{\"o}rn Ommer}.} \bibinfo{year}{2022}\natexlab{}.
\newblock \showarticletitle{High-resolution image synthesis with latent diffusion models}. In \bibinfo{booktitle}{\emph{Proceedings of the IEEE/CVF Conference on Computer Vision and Pattern Recognition}}. \bibinfo{pages}{10684--10695}.
\newblock


\bibitem[Rudin(1986)]%
        {babyrudin}
\bibfield{author}{\bibinfo{person}{W. Rudin}.} \bibinfo{year}{1986}\natexlab{}.
\newblock \bibinfo{booktitle}{\emph{Principles of Mathematical Analysis}}.
\newblock \bibinfo{publisher}{McGraw - Hill Book C.}
\newblock
\urldef\tempurl%
\url{https://books.google.com/books?id=frdNAQAACAAJ}
\showURL{%
\tempurl}


\bibitem[Seabold and Perktold(2010)]%
        {statsmodels}
\bibfield{author}{\bibinfo{person}{Skipper Seabold} {and} \bibinfo{person}{Josef Perktold}.} \bibinfo{year}{2010}\natexlab{}.
\newblock \showarticletitle{statsmodels: Econometric and statistical modeling with python}. In \bibinfo{booktitle}{\emph{9th Python in Science Conference}}.
\newblock


\bibitem[Shorten and Khoshgoftaar(2019)]%
        {shorten2019survey}
\bibfield{author}{\bibinfo{person}{Connor Shorten} {and} \bibinfo{person}{Taghi~M Khoshgoftaar}.} \bibinfo{year}{2019}\natexlab{}.
\newblock \showarticletitle{A survey on image data augmentation for deep learning}.
\newblock \bibinfo{journal}{\emph{Journal of big data}} \bibinfo{volume}{6}, \bibinfo{number}{1} (\bibinfo{year}{2019}), \bibinfo{pages}{1--48}.
\newblock


\bibitem[Simonyan and Zisserman(2015)]%
        {vgg}
\bibfield{author}{\bibinfo{person}{Karen Simonyan} {and} \bibinfo{person}{Andrew Zisserman}.} \bibinfo{year}{2015}\natexlab{}.
\newblock \bibinfo{title}{Very Deep Convolutional Networks for Large-Scale Image Recognition}.
\newblock
\newblock
\showeprint[arxiv]{1409.1556}~[cs.CV]


\bibitem[Szegedy et~al\mbox{.}(2014)]%
        {szegedy2014intriguing}
\bibfield{author}{\bibinfo{person}{Christian Szegedy}, \bibinfo{person}{Wojciech Zaremba}, \bibinfo{person}{Ilya Sutskever}, \bibinfo{person}{Joan Bruna}, \bibinfo{person}{Dumitru Erhan}, \bibinfo{person}{Ian~J. Goodfellow}, {and} \bibinfo{person}{Rob Fergus}.} \bibinfo{year}{2014}\natexlab{}.
\newblock \showarticletitle{Intriguing properties of neural networks}. In \bibinfo{booktitle}{\emph{2nd International Conference on Learning Representations, {ICLR} 2014, Banff, AB, Canada, April 14-16, 2014, Conference Track Proceedings}}, \bibfield{editor}{\bibinfo{person}{Yoshua Bengio} {and} \bibinfo{person}{Yann LeCun}} (Eds.).
\newblock
\urldef\tempurl%
\url{http://arxiv.org/abs/1312.6199}
\showURL{%
\tempurl}


\bibitem[Virtanen et~al\mbox{.}(2020)]%
        {scipy}
\bibfield{author}{\bibinfo{person}{Pauli Virtanen}, \bibinfo{person}{Ralf Gommers}, \bibinfo{person}{Travis~E. Oliphant}, \bibinfo{person}{Matt Haberland}, \bibinfo{person}{Tyler Reddy}, \bibinfo{person}{David Cournapeau}, \bibinfo{person}{Evgeni Burovski}, \bibinfo{person}{Pearu Peterson}, \bibinfo{person}{Warren Weckesser}, \bibinfo{person}{Jonathan Bright}, \bibinfo{person}{St{\'e}fan~J. {van der Walt}}, \bibinfo{person}{Matthew Brett}, \bibinfo{person}{Joshua Wilson}, \bibinfo{person}{K.~Jarrod Millman}, \bibinfo{person}{Nikolay Mayorov}, \bibinfo{person}{Andrew R.~J. Nelson}, \bibinfo{person}{Eric Jones}, \bibinfo{person}{Robert Kern}, \bibinfo{person}{Eric Larson}, \bibinfo{person}{C~J Carey}, \bibinfo{person}{{\.I}lhan Polat}, \bibinfo{person}{Yu Feng}, \bibinfo{person}{Eric~W. Moore}, \bibinfo{person}{Jake {VanderPlas}}, \bibinfo{person}{Denis Laxalde}, \bibinfo{person}{Josef Perktold}, \bibinfo{person}{Robert Cimrman}, \bibinfo{person}{Ian Henriksen}, \bibinfo{person}{E.~A. Quintero},
  \bibinfo{person}{Charles~R. Harris}, \bibinfo{person}{Anne~M. Archibald}, \bibinfo{person}{Ant{\^o}nio~H. Ribeiro}, \bibinfo{person}{Fabian Pedregosa}, \bibinfo{person}{Paul {van Mulbregt}}, {and} \bibinfo{person}{{SciPy 1.0 Contributors}}.} \bibinfo{year}{2020}\natexlab{}.
\newblock \showarticletitle{{{SciPy} 1.0: Fundamental Algorithms for Scientific Computing in Python}}.
\newblock \bibinfo{journal}{\emph{Nature Methods}}  \bibinfo{volume}{17} (\bibinfo{year}{2020}), \bibinfo{pages}{261--272}.
\newblock
\urldef\tempurl%
\url{https://doi.org/10.1038/s41592-019-0686-2}
\showDOI{\tempurl}


\bibitem[Wang et~al\mbox{.}(2020)]%
        {tofu}
\bibfield{author}{\bibinfo{person}{Zi Wang}, \bibinfo{person}{Ben Liblit}, {and} \bibinfo{person}{Thomas~W. Reps}.} \bibinfo{year}{2020}\natexlab{}.
\newblock \showarticletitle{{TOFU:} Target-Orienter FUzzer}.
\newblock \bibinfo{journal}{\emph{CoRR}}  \bibinfo{volume}{abs/2004.14375} (\bibinfo{year}{2020}).
\newblock
\showeprint[arXiv]{2004.14375}
\urldef\tempurl%
\url{https://arxiv.org/abs/2004.14375}
\showURL{%
\tempurl}


\bibitem[Wasserman(2010)]%
        {stat}
\bibfield{author}{\bibinfo{person}{Larry Wasserman}.} \bibinfo{year}{2010}\natexlab{}.
\newblock \bibinfo{booktitle}{\emph{All of statistics : a concise course in statistical inference}}.
\newblock \bibinfo{publisher}{Springer}, \bibinfo{address}{New York}.
\newblock
\showISBNx{9781441923226 1441923225}
\urldef\tempurl%
\url{http://www.amazon.de/All-Statistics-Statistical-Inference-Springer/dp/1441923225/ref=sr_1_2?ie=UTF8&qid=1356099149&sr=8-2}
\showURL{%
\tempurl}


\bibitem[Webber et~al\mbox{.}(2010)]%
        {rbo}
\bibfield{author}{\bibinfo{person}{William Webber}, \bibinfo{person}{Alistair Moffat}, {and} \bibinfo{person}{Justin Zobel}.} \bibinfo{year}{2010}\natexlab{}.
\newblock \showarticletitle{A Similarity Measure for Indefinite Rankings}.
\newblock \bibinfo{journal}{\emph{ACM Trans. Inf. Syst.}} \bibinfo{volume}{28}, \bibinfo{number}{4}, Article \bibinfo{articleno}{20} (\bibinfo{date}{nov} \bibinfo{year}{2010}), \bibinfo{numpages}{38}~pages.
\newblock
\showISSN{1046-8188}
\urldef\tempurl%
\url{https://doi.org/10.1145/1852102.1852106}
\showDOI{\tempurl}


\bibitem[Wen et~al\mbox{.}(2019)]%
        {api-misuse1}
\bibfield{author}{\bibinfo{person}{Ming Wen}, \bibinfo{person}{Yepang Liu}, \bibinfo{person}{Rongxin Wu}, \bibinfo{person}{Xuan Xie}, \bibinfo{person}{Shing-Chi Cheung}, {and} \bibinfo{person}{Zhendong Su}.} \bibinfo{year}{2019}\natexlab{}.
\newblock \showarticletitle{Exposing Library API Misuses Via Mutation Analysis}. In \bibinfo{booktitle}{\emph{2019 IEEE/ACM 41st International Conference on Software Engineering (ICSE)}}. \bibinfo{pages}{866--877}.
\newblock
\urldef\tempurl%
\url{https://doi.org/10.1109/ICSE.2019.00093}
\showDOI{\tempurl}


\bibitem[Xie et~al\mbox{.}(2023)]%
        {diversity-boosting}
\bibfield{author}{\bibinfo{person}{Xiaoyuan Xie}, \bibinfo{person}{Pengbo Yin}, {and} \bibinfo{person}{Songqiang Chen}.} \bibinfo{year}{2023}\natexlab{}.
\newblock \showarticletitle{Boosting the Revealing of Detected Violations in Deep Learning Testing: A Diversity-Guided Method}. In \bibinfo{booktitle}{\emph{Proceedings of the 37th IEEE/ACM International Conference on Automated Software Engineering}} (Rochester, MI, USA) \emph{(\bibinfo{series}{ASE '22})}. \bibinfo{publisher}{Association for Computing Machinery}, \bibinfo{address}{New York, NY, USA}, Article \bibinfo{articleno}{17}, \bibinfo{numpages}{13}~pages.
\newblock
\showISBNx{9781450394758}
\urldef\tempurl%
\url{https://doi.org/10.1145/3551349.3556919}
\showDOI{\tempurl}


\bibitem[Zalewski(2022)]%
        {afl}
\bibfield{author}{\bibinfo{person}{Michal Zalewski}.} \bibinfo{year}{2022}\natexlab{}.
\newblock \bibinfo{title}{american fuzzy lop}.
\newblock \bibinfo{howpublished}{\url{https://lcamtuf.coredump.cx/afl/}}.
\newblock


\end{thebibliography}
